\documentclass[letterpaper,preprint,11pt,fleqn,preprintnumbers,prd,onecolumn,superscriptaddress,longbibliography]{revtex4-2}

\usepackage[bookmarks=true]{hyperref}
\usepackage{graphicx,epsfig,color}
\usepackage[english]{babel}
\usepackage{amssymb,amsfonts,amsmath,mathtools}
\usepackage{verbatim}
\makeatletter\AtBeginDocument{\let\@elt\relax}\makeatother

\newcommand{\be}{\begin{equation}} \newcommand{\ee}{\end{equation}}
\newcommand{\bea}{\begin{eqnarray}} \newcommand{\eea}{\end{eqnarray}}

\newcommand{\OO}{\mathcal{O}}

\def\d{\mathrm{d}}

\newcommand{\vev}[1]{{\left\langle #1 \right\rangle}}

\begin{document}

\title{Effective actions and bubble nucleation from holography}

\author{F\"eanor Reuben Ares}
\email{F.R.Ares@sussex.ac.uk}
\affiliation{Department of Physics and Helsinki Institute of Physics\\
P.O.~Box 64, FI-00014 University of Helsinki, Finland}
\affiliation{Department of Physics \& Astronomy, University of Sussex\\
Brighton, BN1 9QH, United Kingdom}

\author{Oscar Henriksson}
\email{oscar.henriksson@helsinki.fi}
\affiliation{Department of Physics and Helsinki Institute of Physics\\
P.O.~Box 64, FI-00014 University of Helsinki, Finland}

\author{Mark Hindmarsh}
\email{mark.hindmarsh@helsinki.fi}
\affiliation{Department of Physics and Helsinki Institute of Physics\\
P.O.~Box 64, FI-00014 University of Helsinki, Finland}
\affiliation{Department of Physics \& Astronomy, University of Sussex\\
Brighton, BN1 9QH, United Kingdom}

\author{Carlos Hoyos}
\email{hoyoscarlos@uniovi.es}
\affiliation{Department of Physics and \\
Instituto de Ciencias y Tecnolog\'{\i}as Espaciales de Asturias (ICTEA),\\
Universidad de Oviedo, c/ Federico Garc\'{\i}a Lorca 18, ES-33007 Oviedo, Spain}

\author{Niko Jokela}
\email{niko.jokela@helsinki.fi}
\affiliation{Department of Physics and Helsinki Institute of Physics\\
P.O.~Box 64, FI-00014 University of Helsinki, Finland}

\begin{abstract}
We discuss the computation of the quantum effective action of strongly interacting field theories using holographic duality, and its use to determine quasi-equilibrium parameters of first order phase transitions relevant for gravitational wave production. A particularly simple holographic model is introduced, containing only the metric and a free massive scalar field. Despite the simplicity, the model contains a rich phase diagram, including first order phase transitions at non-zero temperature, due to various multi-trace deformations. We obtain the leading terms in the effective action from homogeneous black brane solutions in the gravity dual, and linearised perturbations around them. We then employ the effective action to construct bubble and domain wall solutions in the field theory side and study their properties. In particular, we show how the scaling of the effective action with the effective number of degrees of freedom of the quantum field theory determines the corresponding scaling of gravitational wave parameters.

\end{abstract}

\preprint{HIP-2021-30/TH}

\maketitle

\newpage

\tableofcontents

\hrulefill
\vspace{10pt}
\newpage

\section{Introduction}

First order phase transitions are of great interest especially for early-universe cosmology, where bubble nucleation could result in the production of an observable gravitational wave signal at LISA \cite{LISA:2017pwj,Caprini:2019egz}, providing evidence for beyond the Standard Model physics. If this new physics is strongly coupled, computations of the parameters of the phase transition relevant for gravitational wave production with 
standard techniques (see, {\emph{e.g.}}, \cite{Mazumdar:2018dfl,Hindmarsh:2020hop} for reviews) fail, and the direct connection between the gravitational wave signature and the masses and couplings of the underlying field theory is lost.

Gravitational waves at strong coupling in SU($N$) gauge theories have been investigated using phenomenological models of the free energy density \cite{Halverson:2020xpg,Huang:2020crf,Reichert:2021cvs}. 
A complementary approach is through holographic duality 
\cite{Bigazzi:2020avc,Ares:2020lbt,Zhu:2021vkj}, but analyses have so far had certain limitations. In \cite{Ares:2020lbt,Zhu:2021vkj} only equilibrium properties were computed, with some significant properties such as the bubble nucleation rates left as free parameters. This was not the case in \cite{Bigazzi:2020avc}, where the transition rate was computed (based on the results in \cite{Bigazzi:2020phm}), but some additional assumptions were made, either by working in a quenched sector of the theory or by using a phenomenological approach that does not strictly follow from the holographic dictionary. In this work we will try to partially improve the holographic approach and present a derivation of the transition rate that does not require these assumptions. Although our analysis is motivated by its possible application to cosmological transitions and thus limited to high temperature and zero charge density, it can be straightforwardly generalised to other set-ups.

In quantum field theories the dynamical evolution of the phase transition can start to be addressed by finding the quantum effective action of the theory, which in many aspects is reminiscent of a Ginzburg-Landau effective action. Bubble configurations are obtained from semiclassical solutions to the effective action, and these can be employed to compute key properties of the transition such as the nucleation rate. Bubble production could take place through quantum tunnelling or thermal fluctuations, the probabilities of which can be estimated from the action. However, the dynamical evolution of the bubbles themselves require further analysis, as at nonzero temperature dissipation and drag will enter into play. We will not attempt to describe the dynamical evolution of bubbles, but this has been studied in some models \cite{Bea:2021zsu,Bigazzi:2021fmq,Henriksson:2021zei}.

In a weakly coupled theory, the effective action can be computed by standard perturbative methods (see, however, \cite{Gould:2021oba} for a discussion of issues at non-zero temperature). At strong coupling things are as always more difficult. While lattice simulations provide one route of attack (see, {\emph{e.g.}}, \cite{Moore:2000jw}), they are computationally expensive and have great difficulty at non-zero charge density and especially for real time evolution. A possible avenue is to use gauge/gravity duality, or holography, which is well suited both for strong coupling and to study the dynamical evolution of the system at non-zero  temperature. 

In this paper, we consider the computation of the effective action using holographic duality. This allows us to study a strongly coupled QFT (often a gauge theory in the large-$N$ limit) through the lens of a classical gravitational theory. We will consider only zero charge density, so our focus is on configurations that may be relevant for a cosmological phase transition, and pick a simple model to illustrate our approach. An analysis of the gravitational wave signal extracted in this model will be presented elsewhere \cite{shortcompanion}. 

Our approach has some similarities with the effective action approach used to describe the confinement-deconfinement transition from holographic models \cite{Bigazzi:2020phm,Janik:2021jbq}, in that we do not attempt to find gravity solutions dual to bubble configurations, but construct the bubble solutions directly in the field theory. However, we do not make any additional phenomenological assumptions within the holographic model. Our derivation of the effective action follows directly from the usual rules of the duality. We will truncate the effective action by keeping only terms with two derivatives, but we show that higher derivative terms seem to be comparatively suppressed in the bubble configurations we obtain.

The outline of the paper is as follows. In Sec.~\ref{sec:effActionHolography} we warm up with a general discussion of the field theory (quantum) effective action, and show how to extract it from the gravity dual. Then, in Sec.~\ref{sec:simpleExample}, we introduce a particularly simple ``bottom-up'' gravity theory which nonetheless displays an interesting phase structure upon deforming it by single- and multi-trace operators. By finding a one-parameter family of numerical black brane solutions and applying careful holographic renormalisation, we show how to extract the effective potential, and thereby produce the phase diagram. Furthermore, by solving the linearised equations of motion around the black brane solutions, we show how to derive the (non-canonical) kinetic term as well as a subset of higher derivative terms. In Sec.~\ref{sec:bubbles} we then use the effective action to study the first order phase transitions of this theory, by finding the critical bubble solutions and computing their action, which sets the nucleation rate. We discuss implications for early-universe cosmology, including the computations of the nucleation temperature and the transition rate, as well as their dependence on the number of degrees of freedom $N$. We also briefly discuss domain walls and compute their surface tension for the complete parameter space, allowing us to comment on the applicability of the  thin-wall approximation. Our conclusions and the discussion of the extensions of our work appear in Sec.~\ref{sec:discussion}. The appendices detail on the holographic renormalisation, exact results at large temperatures, and also the linearised fluctuation equations.

\section{The quantum effective action from holography}\label{sec:effActionHolography}

Consider a theory with a scalar field $\Psi$ whose action we denote $S[\Psi]$. The path integral in the presence of an external source $J$ is
\begin{equation}
 {\cal Z}[J] = \int \mathcal{D}\Psi\, \exp\left[ iS[\Psi] + i\int d^4x J\Psi \right] \ .
\end{equation}
From the path integral one can obtain the closely related generating functional for \emph{connected} correlation functions
\begin{equation}
 {\cal W}[J] = -i \log{{\cal Z}[J]} \ ,
\end{equation}
we define the effective action through a functional Legendre transform,
\begin{equation}\label{eq:effActionDef}
 \Gamma[\vev{\Psi}_J] = {\cal W}[J] - \int d^4 x\, \vev{\Psi}_J J \ .
\end{equation}
In this definition $J$ should be understood as being a functional of $\vev{\Psi}_J$ determined implicitly through the relationship
\begin{equation}
 \frac{\delta {\cal W}[J]}{\delta J} = \vev{\Psi}_J \ .
\end{equation}
This is also the statement that $\vev{\Psi}_J $ --- sometimes referred to as the classical field --- corresponds to the expectation value of $\Psi$ for a given source $J$. Separating the expectation value of the field in the sourceless and sourced parts
\begin{equation}
\psi=\vev{\Psi}_{J=0},\ \ \delta\psi=\vev{\Psi}_J-\vev{\Psi}_{J=0} \ ,
\end{equation}
the effective action can be recast as a functional of $\delta\psi$. The effective action so defined is the generating functional of 1-point irreducible (1PI) connected correlation functions $\Gamma_n(x_1,\ldots,x_n;\psi)$; hence it can be expanded as
\begin{equation}\label{eq:gammaexp}
 \Gamma[\psi+\delta\psi] = \sum_{n=0}^\infty \frac{1}{n!}\int d^4x_1\ldots d^4x_n \Gamma_n(x_1,\ldots,x_n;\psi)\delta\psi(x_1) \ldots \delta\psi(x_n) \ .
\end{equation}
The 1PI connected correlators themselves admit in principle an expansion around the trivial vacuum $\psi=0$
\begin{equation}
\begin{split}
\Gamma_n(x_1,\ldots,x_n;\psi)&=\Gamma_n(x_1,\ldots,x_n;0)\\
&+ \sum_{k\geq 1} \frac{1}{k!}\int d^4y_1\cdots d^4y_k \Gamma_{n+k}(x_1,\ldots,x_n,y_1,\ldots,y_k;0)\psi(y_1)\cdots \psi(y_k) \ . 
\end{split}
\end{equation}	

The full effective action is in general highly non-local, as can be seen from this expansion. However, if there are no gapless degrees of freedom, at sufficiently low energies one expects that it can be written as an integral over a local effective Lagrangian, and furthermore, that it can be expanded for small derivatives:
\begin{equation}
 \Gamma[\psi+\delta\psi] = \int d^4x \left[ -V(\psi+\delta\psi) - \frac{1}{2}Z(\psi+\delta\psi)\partial_\mu \delta\psi \, \partial^\mu \delta\psi + \ldots \right] \ .
\end{equation}
Here $V(\psi+\delta\psi)$ and $Z(\psi+\delta\psi)$ are ordinary functions of $\psi+\delta\psi$, and we have assumed that the sourceless state is static and homogeneous at $J=0$, so that $\partial_\mu\psi=0$. The function $V(\psi+\delta\psi)$ is known as the effective potential, whose minimum determines the true ground state of the theory.

If $\psi$ coincides with the ground state, further expanding to quadratic order in $\delta\psi$ leads to
\begin{equation}\label{eq:derivativeExpansion}
 \Gamma[\psi+\delta\psi] = \int d^4x \left[ -V(\psi)-\frac{1}{2}V''(\psi)(\delta\psi)^2 - \frac{1}{2}Z(\psi)\partial_\mu \delta\psi \, \partial^\mu \delta\psi + \ldots \right] \ .
\end{equation}
In order to extract the coefficients in the effective action we compare the expansions in \eqref{eq:gammaexp} and \eqref{eq:derivativeExpansion}. Going to momentum space and expanding the correlators $\Gamma_2$ around zero frequency and vanishing spatial momenta (we omit the dependence on $\psi$ and factor out a Dirac delta imposing momentum conservation),
\begin{equation}
 \widetilde{\Gamma}_2(k) = \widetilde{\Gamma}_2(0) + \frac{1}{2}\frac{\partial^2 \widetilde{\Gamma}_2}{\partial k_i \partial k_j}\Big|_{k=0} k_i k_j + \ldots\;,
\end{equation}
one can deduce that 
\begin{equation}\label{eq:Zexpansion}
V''(\psi)=-\widetilde{\Gamma}_2(0),\ \ Z(\psi)=-\frac{1}{6}\delta_{ij}\frac{\partial^2 \widetilde{\Gamma}_2}{\partial k_i \partial k_j}\Big|_{k=0} \ .
\end{equation}

We will be interested in constructing the effective action up to (at least) second order in the derivative expansion; that is, we want to compute $V(\psi)$ and $Z(\psi)$. We do this in the framework of holographic duality, which lets us study a strongly coupled quantum field theory by solving a classical gravitational one. The essential relationship in the holographic dictionary is the equivalence between the renormalised on-shell gravitational action and the field theory generating functional ${\cal W}[J]$. Thus, if one can come by a set of solutions to the gravitational field equations corresponding to different sources $J$ (meaning different near-boundary falloffs for the fields of interest) one can simply evaluate the gravitational action on these solutions to find ${\cal W}[J]$, and then Legendre transform to obtain $\Gamma[\vev{\Psi}_J]$. 

Of course, even the solution of classical field equations for arbitrary boundary conditions can be extremely challenging. However, assuming unbroken translational symmetry in the field theory directions one can typically find such solutions numerically for a large class of theories, including the simple gravity plus scalar theories we focus on later in the paper. This lets us compute $\Gamma[\psi]$ in the limit of uniform fields and sources, {\emph{i.e.}}, the effective potential.

To access derivative terms in the effective action, we then perturb away from these uniform solutions. The essential insight is that from (\ref{eq:Zexpansion}), the value of $Z(\psi)$ at some particular value of $\psi$ is just given by the leading terms in the low momentum expansion of the two-point correlators. This is readily done in holography by solving the gravitational field equations linearised around a particular background solution.

Note that since we will be interested in the effective action at some non-zero temperature, the Lorentz invariance displayed in for example (\ref{eq:derivativeExpansion}) will be broken. We will mainly be interested in static field configurations, and so limit ourselves to computing the coefficient of the spatial derivatives. Generalizing by including derivatives with respect to time is straightforward.

Previous authors have discussed computing the field theory effective action through holography \cite{Hertog:2004ns,Hertog:2005hu,Papadimitriou:2007sj,Kiritsis:2012ma,Kiritsis:2014kua}. Of these, several make use of (fake) superpotential formulations on the gravity side to derive analytical expressions for the effective action. In simplifying limits such as close to conformality these are very useful. In order to describe thermal phase transitions, as our aim is here, further (numerical) work is typically needed. Our approach is in some sense more direct, employing numerics from the outset; the effective potential analysis in references \cite{Hertog:2004ns,Hertog:2005hu} are the closest in spirit.

\subsection{Holographic duality}\label{sec:holography}

We now give a brief introduction to holographic duality, and argue that the approach we outlined for computing the effective action in a derivative expansion is quite natural and convenient in this setting.  

Holographic duality relates a $d$-dimensional quantum field theory (QFT) with a $D=(d+1)$-dimensional gravitational theory in an (asymptotically) anti-de Sitter (AdS) spacetime of radius $L$. The AdS behaviour corresponds to a fixed point in the UV of the QFT. If the space is AdS throughout, then the dual is a conformal field theory (CFT), while asymptotically AdS spacetimes correspond to perturbations of the fixed point by some relevant operators. Well-understood examples of this duality originate in string theory, where the QFT is typically a gauge theory with some amount of supersymmetry. To be able to suppress quantum and string effects in the bulk, rendering the gravitational theory classical, one must typically take the limit of many degrees of freedom and strong coupling. If the dual QFT is a gauge theory, the former can be realised as a large-$N$ limit where $N$ is the rank of the group. In CFTs the number of degrees of freedom can be associated with the central charge (in two dimensions), conformal anomaly coefficients (in even dimensions) or other quantities. We will refer to $N$ as the ``number of colours'', even if the dual field theory is not known.

In slightly more detail, the gravitational constant $\propto\kappa_5^2$ (whose inverse multiplies the 5D gravity action), made dimensionless by dividing by the appropriate power of the radius of curvature $L$, is related to the number of degrees of freedom. When the dual field theory is a rank-$N$ gauge theory, in particular, we typically have a relation of the form
\begin{equation}
 \frac{L^3}{\kappa^2_5} \propto N^2 \ .
\end{equation}
Since we work in a bottom-up setting, the detailed form of this relationship is not known. For simplicity, we will set $L^3/\kappa^2_5 = N^2$, treating $N$ as a free parameter related to the number of degrees of freedom, while keeping in mind it is not necessarily equal to the rank of some gauge group.

The field theory effective action we compute through holography will also have this large pre-factor $N^2$. This is important for bubble nucleation, since a large bubble action exponentially suppresses the nucleation rate. We will discuss this issue in detail in Sec.~\ref{sec:bubbles}. For now, we only note that we will explicitly add this factor of $N^2$ in (\ref{eq:derivativeExpansion}), writing it as
\begin{equation}\label{eq:derivativeExpansionN2}
 \Gamma[\psi+\delta\psi] = N^2 \int d^4x \left[ -V(\psi)-\frac{1}{2}V''(\psi)(\delta\psi)^2 - \frac{1}{2}Z(\psi)\partial_\mu \delta\psi \, \partial^\mu \delta\psi + \ldots \right] \ .
\end{equation}
Thus, in the rest of the paper the quantities $V(\psi)$ and $Z(\psi)$ are $\mathcal{O}(1)$, while the full effective action is $\mathcal{O}(N^2)$.

Through holographic duality, the operator $\Psi$ is associated with a scalar field $\phi$ in a gravitational theory. The gravitational theory admits classical solutions which are asymptotically anti-de Sitter --- near the boundary of these spacetimes, the metric approaches the form
\begin{equation}
ds^2 = \frac{r^2}{L^2}\eta_{\mu\nu}dx^\mu dx^\nu+\frac{L^2}{r^2}dr^2\ \ ,\ \ r\to \infty \ ,
\end{equation}
and the field $\phi$ will fall off as
\begin{equation}
 \phi(r,x) = \frac{\phi_-(x)}{r^{\Delta_-}} + \frac{\phi_+(x)}{r^{\Delta_+}} + \ldots \ ,
\end{equation}
where $\Delta_\pm=\frac{d}{2}\pm \sqrt{\frac{d^2}{4}+m^2L^2}$. 

In a typical realisation of holographic duality, the operators dual to classical fields on the gravity side will be some form of gauge-invariant single-trace operators, meaning they contain a single trace over colour indices. A deformation of the theory consisting of introducing a source for a scalar single-trace operator like $\Psi$ is realised in the gravity dual by imposing a boundary condition such that $\phi_-$ is non-zero. One can also study deformations by operators with two or more traces; due to large-$N$ factorisation, such operators have a simple description on the gravity side. These will prove useful for us in the next section, as they provide ``knobs'' to turn in order to make our simple gravity dual exhibit first order phase transitions.

For relevant multi-trace deformations to be possible, the mass of the scalar field must be close to the Breitenlohner-Freedman bound, in the range
\begin{equation}\label{eq:MTrange}
 -\left(\frac{d}{2}\right)^2 \le m^2L^2 \le -\left(\frac{d}{2}\right)^2 + 1 \ .
\end{equation}
In this range, the value of the mass allows for alternate quantisation, where the coefficient of the sub-leading falloff $\phi_+$ is fixed and identified as the source $J$ of the dual operator. In this case the leading falloff $\phi_-$ is proportional to the expectation value $\vev{\Psi}_J$ of the operator.

Once in alternate quantisation, multi-trace deformations are implemented by generalising the boundary condition on the scalar field \cite{Witten:2001ua}, allowing $\phi_+$ to be given by an arbitrary function of $\phi_-$. More specifically, if we want to deform our theory by some general multi-trace deformation $W(\Psi)$, we should impose the boundary condition
\begin{equation}
 \phi_+ = \frac{\delta W(\vev{\Psi})}{\delta \vev{\Psi}}  \ ,
\end{equation}
where we recall that $\vev{\Psi}$ is proportional to $\phi_-$. For example, in the theory we discuss in the next section, we show through careful holographic renormalisation that $\vev{\Psi}=-4\phi_-/3$. Then, deforming by say a double trace deformation $W(\Psi)=f\Psi^2/2$ means imposing the boundary condition
\begin{equation}
 \phi_+ = f\vev{\Psi} = -\frac{4}{3}f\, \phi_- \ .
\end{equation}

The multi-trace deformations we implement have a straightforward effect on the field theory effective action; a deformation by an $n$-trace operator $\Psi^n$ simply adds a term $\propto \psi^n$ to the effective potential. For single-trace deformations ($n=1$) this is in fact a general result for all field theories, following from the behaviour of the Legendre transform under a shift. The fact that multi-trace deformations leads to simple polynomial contributions is on the other hand only true in the large-$N$ limit (see, {\emph{e.g.}}, \cite{Papadimitriou:2007sj}).

\section{Concrete example: CFT with a dimension-4/3 operator}\label{sec:simpleExample}

We now apply the general ideas from the previous section to study a remarkably simple gravitational theory whose field theory dual enjoys a first order phase transition at non-zero temperature for a range of parameters. We work in a bottom-up setting, meaning we select a simple gravity theory capturing the features we are interested in. In this case, besides gravity with a negative cosmological constant, all we need is a scalar operator which can act as order parameter for the phase transition. We thus take the gravitational action to be
\begin{equation}\label{eq:gravityAction}
 S_{bulk}=\frac{1}{2\kappa^2_5}\int d^{5}x\sqrt{-g}\left[\mathcal{R}-\partial_\mu\phi\partial^\mu\phi-\mathcal{P}(\phi)\right] \ .
\end{equation}
Here $\mathcal{R}$ is the scalar curvature, $g$ is the metric of the five-dimensional asymptotically anti-de Sitter (AdS) spacetime, $\phi$ is the scalar field, and $\kappa_5^2=8\pi G_5$ is essentially the Newton constant.
The potential for the scalar field $\phi$ is taken to have the minimal form
\begin{equation}
\mathcal{P}(\phi)=-\frac{12}{L^2}+m^{2}\phi^{2}\ ,
\end{equation}
with $m^2L^2=-32/9$ \footnote{The paper \cite{Papadimitriou:2007sj} discusses a class of potentials, dubbed the '2/3' potential, which has the same mass plus higher order terms; with four instead of five bulk dimensions, this potential can be embedded in $\mathcal{N}=8$ gauged supergravity.}. From this point on we set the radius of curvature $L=1$. The value of $m^2$ is within the range of masses allowing two possible quantisations --- we will select the alternate quantisation, meaning that the dual operator has dimension $\Delta=4/3$. Choosing alternate quantisation allows for relevant multi-trace deformations, which provides us with useful ``knobs'' to turn (in addition to a single-trace deformation and temperature) to arrive at a theory with a first order (thermal) phase transition. We will consider deformations of the original dual CFT, with a scalar operator $\Psi$ and action $S_{CFT}$, by single-, double-, and triple-trace deformations,
\begin{equation}
 S_{CFT}\rightarrow S_{CFT}+\int\d^4x \left( \Lambda\Psi+\frac{f}{2}\Psi^{2}+\frac{g}{3}\Psi^{3} \right) \ .
\end{equation}
The choice of $\Delta=4/3$ for the operator is convenient as it means that the triple trace deformation is marginal. Thus, the coupling $g$ is dimensionless, while $\Lambda$ and $f$ have dimensions $8/3$ and $4/3$, respectively.

\subsection{Finding background solutions}\label{sec:findingBackgrounds}

Since we want to study the field theory at non-zero temperature, we search for black brane solutions of the gravitational theory. A convenient Ansatz is
\begin{equation}
ds^{2}=-e^{-2\chi(r)}h(r)dt^{2}+\frac{{dr^{2}}}{h(r)}+r^{2}d\vec{x}^{2}\,,
\end{equation}
and $\phi=\phi(r)$. The equations of motion (EoM) for our system can then be written as follows:
\begin{gather}
\chi'(r)+\frac{r}{3}\phi'(r)^{2}=0 \label{eq:EoM1}\\
h'(r)+h(r)\left(\frac{{2}}{r}+\frac{{r}}{3}\phi'(r)^{2}\right)+\frac{{r}}{3}\mathcal{P}(\phi(r))=0 \label{eq:EoM2}\\
\phi''(r)+\frac{{\phi'(r)}}{r}-\frac{{2r\mathcal{P}(\phi(r))\phi'(r)+3\mathcal{P}'(\phi(r))}}{6h(r)}=0 \label{eq:EoM3} \ .
\end{gather}
The equations allow for an AdS solution. Near the boundary of \emph{asymptotically} AdS solutions, the fields fall off as
\begin{align}\label{eq:falloff}
\phi& = \frac{\phi_-}{r^{4/3}}+\frac{\phi_+}{r^{8/3}}+\ldots \nonumber \\
h& =  r^{2}+\frac{{4}}{9}\frac{{\phi_-^{2}}}{r^{2/3}}+\frac{{h_{2}}}{r^{2}}+\ldots \\
\chi& = \chi_{0}+\frac{{2}}{9}\frac{\phi_-^{2}}{r^{8/3}}+\ldots \ . \nonumber
\end{align}
We use standard numerical methods, implemented using \textit{Mathematica}'s \textbf{NDSolve} function, to look for hairy black brane solutions. We begin by noting that (\ref{eq:EoM2}) and (\ref{eq:EoM3}) involve only $h(r)$ and $\phi(r)$ (and not $\chi(r)$). We solve these two equations by expanding them in a power series near the black brane horizon at $r=r_H$, imposing that $h(r)$ goes to zero there and that $\phi(r)$ is regular. The resulting near-horizon series solution has two parameters that are not fixed by the equations of motion; the horizon radius $r_H$ and the value of the scalar at the horizon $\phi(r_{H})\equiv\phi_H$. For each choice of $r_H$ and $\phi_H$, we can numerically integrate the two equations from the horizon to the AdS boundary to obtain a solution for $h(r)$ and $\phi(r)$. We then plug the solution for $\phi(r)$ into (\ref{eq:EoM1}) and solve it for $\chi(r)$, imposing $\chi_{0}=0$ to recover the standard AdS metric near the boundary.

Resulting black brane solutions will be dual to the field theory at non-zero temperature. The temperature and entropy density are set by the Hawking temperature and horizon area of the black brane, given by
\begin{equation}\label{eq:thermo}
 T = \frac{e^{-\chi(r_H)}h'(r_H)}{4\pi} \qquad \text{and} \qquad s = \frac{r_H^3}{4G_5} \ ,
\end{equation}
where we remind that $G_5$ is related to $\kappa_5$ appearing in the pre-factor of the action (\ref{eq:gravityAction}) by $2\kappa^2_5=16\pi G_5$. We want to construct the field theory effective potential at \emph{fixed temperature}. Using the near-horizon expansion, the above expression for the temperature can be seen to take the form
\begin{equation}
 T = - \frac{e^{\chi(r_H)} \mathcal{P}(\phi_H)r_H}{12\pi} \ .
\end{equation}
We see that given one of the free parameters, say $\phi_H$, we can tune the other one, $r_H$, to set the temperature. We use this to set $T=1$ for all our black brane solutions --- all the results we give will thus be in units of temperature. Note that if we had fixed $r_H$ in some other way, say by setting $r_H=1$ (fixing the entropy), we would have to rescale the solutions such that they all have the same temperature before Legendre transforming to get the effective potential.

\begin{figure}
 \includegraphics[scale=0.8]{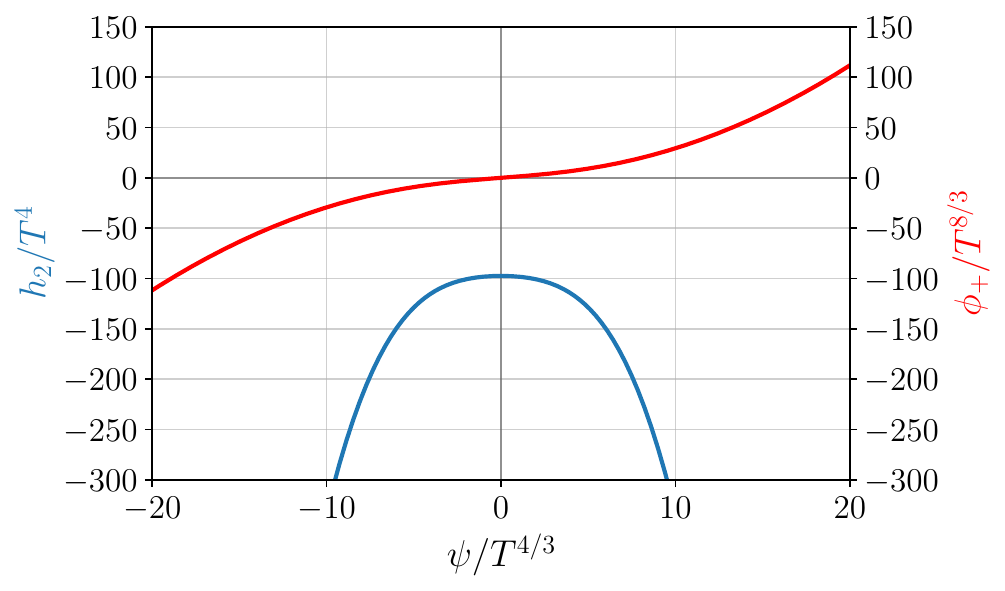}
 \caption{The coefficients $h_{2}$ (lower blue curve) and $\phi_+$ (upper red curve) from (\ref{eq:falloff}) as functions of $\psi=-\frac{4}{3}\phi_-$, all in units of temperature.\label{fig:BHfalloffs}}
\end{figure}

Having fixed $r_H$ we are left with a one-parameter family of solutions, one for each $\phi_H$. It is useful for our purposes to parameterise the solutions by the quantity $\psi=-\frac{4}{3}\phi_-$, which as we show in Appendix~\ref{app:holoRenorm} equals the expectation value of the scalar operator in the dual field theory. We visualise our family of solutions by plotting the coefficients $h_2$ and $\phi_+$ from (\ref{eq:falloff}) as functions of $\psi$ in Fig. \ref{fig:BHfalloffs}.

\subsection{Effective potential}

There are two, essentially equivalent, ways of constructing the field theory effective potential. From (\ref{eq:effActionDef}) and (\ref{eq:derivativeExpansionN2}) we see that for uniform fields and sources, the effective potential can be written as
\begin{equation}\label{eq:effPotUniform}
 V(\psi) = -w(J)+\psi J \ ,
\end{equation}
where we defined $w(J)={\cal W}[J]/(\beta V_3 N^2)$ with $V_3$ being the volume along the spatial directions of the dual field theory, and $\beta=1/T$ being the extent of the Euclidean time direction. Note that we have also implicitly redefined $\psi$ by a factor of $N^2=\kappa^{-2}_5$ as compared with section \ref{sec:effActionHolography}, as is also done in the appendix, see (\ref{eq:EValtQuant}). From the holographic dictionary, we have $\beta V_3 N^2 w(J)=S_{OS}$, where $S_{OS}$ is the full on-shell gravitational action (including all counter-terms). In Appendix~\ref{app:holoRenorm} we go through the holographic renormalisation for our theory, which gives us expressions for $w$, $\psi$ and $J$ in terms of the coefficients in the asymptotic expansion (\ref{eq:falloff}). The end result is \eqref{eq:effPotCompleteA}, which we reproduce here for convenience:
\begin{equation}\label{eq:effPotComplete}
 V(\psi) = \frac{h_{2}(\psi)}{2} + \frac{7}{9}\psi\, \phi_+(\psi) + \Lambda \psi + \frac{f}{2}\psi^2 + \frac{g}{3}\psi^3 \ .
\end{equation}
From our family of black brane solutions we can extract the functions $h_{2}(\psi)$ and $\phi_+(\psi)$ (which are plotted in Fig.~\ref{fig:BHfalloffs}), allowing us to evaluate $V(\psi)$. Note that as mentioned in the previous section, the addition of single-, double-, and triple-trace deformations give linear, quadratic, and cubic contributions to the effective potential, respectively; this should be true in general for a holographic theory in the classical gravity limit.

As an alternative road to the effective potential, introduced in \cite{Hertog:2004ns}, we note that (still assuming uniform fields and sources)
\begin{equation}\label{eq:effPotDerivative}
 \frac{dV(\psi)}{d\psi} = J(\psi) \ .
\end{equation}
Since we can easily extract the curve $J(\psi)$ from our family of black brane solutions using the results obtained from holographic renormalisation in Appendix \ref{app:holoRenorm}, we can simply integrate it (numerically) to obtain $V(\psi)$ up to a constant. And this constant can in fact be fixed by noting that the effective potential at small $\psi$ corresponds to the free energy of pure AdS-Schwarzschild, which is easily obtained. We have checked that these two approaches to compute the effective potential agree.

\begin{figure}
 \includegraphics[scale=0.8]{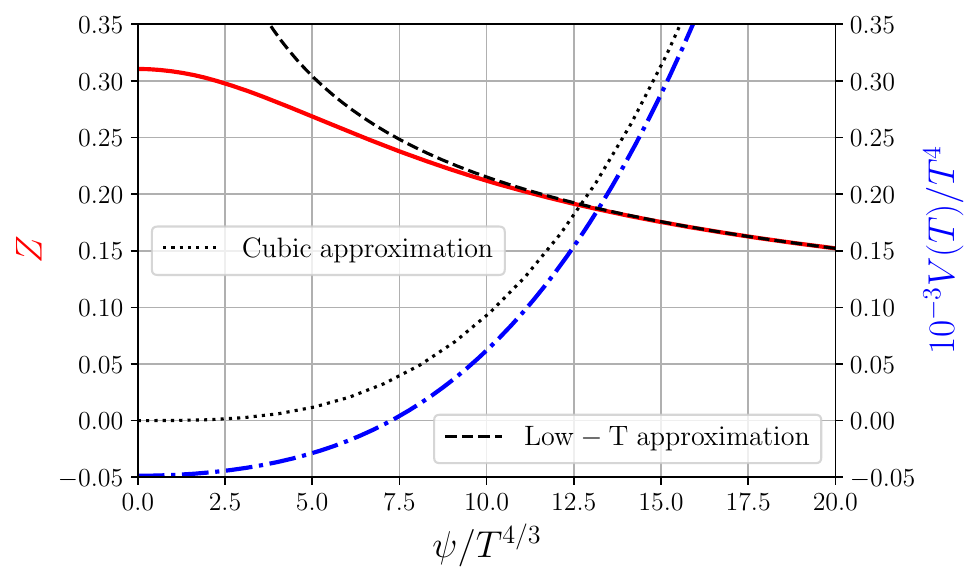}
 \caption{The quantum effective potential of the undeformed theory (dashed-dotted blue), and the kinetic term (solid red, discussed in the next subsection).}\label{fig:effPot_undeformed}
\end{figure}

For the $\Lambda=f=g=0$ theory, which we refer to as the \emph{undeformed} case, this procedure gives us the dashed-dotted blue curve in Fig.~\ref{fig:effPot_undeformed}. As one might have expected from such a simple gravity dual there are no exciting features, only a convex potential with a single minimum. Since the gravity theory is symmetric under $\phi\rightarrow-\phi$, $V(\psi)$ is an even function. At small field values it goes as
\begin{equation}\label{eq:smallPsiExpansion}
 V(\psi) = V_0 + \frac{V_2}{2} \psi^2 + \OO(\psi^4) \ .
\end{equation}
As small $\psi$ is equivalent to large temperatures, in this limit the background approaches AdS-Schwarzschild. Then, $V_0$ can be seen to simply equal the free energy density of this solution,
\begin{equation}
 V_0 = f_{AdS-Sch} = -\frac{\pi^4}{2} \approx -48.70 \ . 
\end{equation}
Moreover, as we show in Appendix \ref{app:effPotV2}, the coefficient $V_2$ can be found exactly by computing the scalar two-point function in an AdS-Schwarzschild spacetime; the result being
\begin{equation}\label{eq:V2}
 V_2 = \frac{9\pi^{17/6}}{\Gamma(1/6)^3} \approx1.337 \ ,
\end{equation}
matching our numerical results well. The explicit temperature dependence of the coefficients in the effective potential follows from simple dimensional analysis: $V_0\sim T^4$ and $V_2\sim T^{4/3}$. Since $V_2>0$, increasing the temperature will tend to stabilise the trivial vacuum $\psi=0$. Our strategy will be to introduce additional terms that destabilise the trivial vacuum at zero temperature, in such a way that we can produce a phase transition when $V_2$ becomes dominant and the trivial vacuum becomes the favoured state at high temperature. 

At large field values or small temperatures, the effective potential grows as
\begin{equation}
 V(\psi) \sim \frac{\gamma_3}{3} |\psi|^{3} \qquad \text{with} \qquad \gamma_3\approx0.278 \ ,
\end{equation}
as shown by the dotted black line in Fig.~\ref{fig:effPot_undeformed}. The cubic behaviour is dictated by the scale invariance of the theory at zero temperature. The full potential cannot be well-fitted by a simple polynomial. 

We now consider deforming the theory by single-, double-, and triple-trace deformations. As discussed in Section \ref{sec:holography}, this will add to the ``undeformed'' effective potential a term linear, quadratic, or cubic in $\psi$, respectively. We want to consider fixing the theory, {\emph{i.e.}}, fixing all couplings, and then tuning the temperature in order to look for a thermal phase transition. Of course, by ``tuning temperature'' we really mean tuning some dimensionless ratio of temperature to some other scale; in our case, it is convenient to define
\begin{equation}
 \tilde T = \frac{T}{|\Lambda|^{3/8}+|f|^{3/4}} \ ,
\end{equation}
which remains well-defined when $\Lambda$ or $f$ (but not both) are zero. In particular, we can just as well tune $\tilde T$ by keeping $T$ fixed and tuning $\Lambda$ and $f$ together (keeping their dimensionless ratio fixed), which is more convenient from a holographic point of view.

Let us emphasise that all the non-trivial strongly coupled physics is in some sense contained in the undeformed effective potential of Fig.~\ref{fig:effPot_undeformed}, which is given by the numerically determined functions $h_2(\psi)$ and $\phi_+(\psi)$ in (\ref{eq:effPotComplete}). The polynomial contributions from single- and multi-trace deformations are on their own reminiscent of a weakly coupled effective description, albeit with the notable difference that the ``order parameter'' $\psi$ has dimension 4/3, meaning the usual $\psi^4$-term is irrelevant. We now describe the effective potential and possible phase transitions that result from turning on different combinations of couplings.

\paragraph{Single-trace deformation ($f=g=0$)}

Adding a single-trace deformation simply shifts the minimum of the undeformed potential around while still keeping the potential convex, not leading to any phase transition or other interesting features.

\paragraph{Double-trace deformation ($\Lambda=g=0$)}

Adding a double-trace deformation is more interesting; for $f<-V_2$, it destabilises the vacuum at $\psi=0$, leading to a double-well potential. The resulting theory thus has a second order thermal phase transition, breaking the $\psi\leftrightarrow-\psi$ symmetry as $\tilde T$ crosses the critical value $V_2^{-3/4}$ from above. This case was studied in \cite{Faulkner:2010gj}.

\paragraph{Triple-trace deformation ($\Lambda=f=0$)}

\begin{figure}
 \includegraphics[scale=0.85]{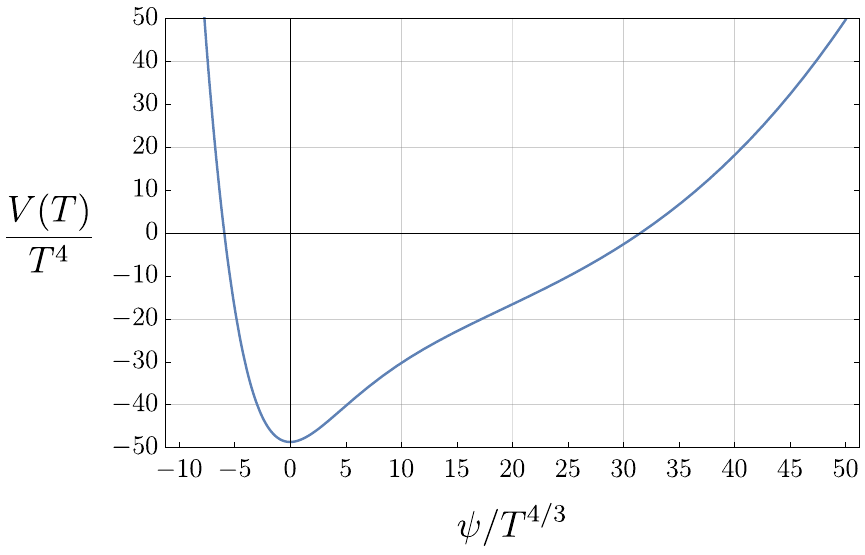}
 \caption{The quantum effective potential of the theory with a triple trace deformation, $g=0.276$.}\label{fig:effPot_tripleTrace0p092}
\end{figure}

Importantly, for large triple-trace deformations $g>\gamma_3$, the potential becomes unbounded from below. Staying below this value, we note that the resulting potential is no longer convex (but still bounded) within the narrow range
\begin{equation}\label{eq:gInt}
 0.2675 \lesssim  g < \gamma_3 \approx 0.278 \ .
\end{equation}
In particular $g=0.276$ results in Fig.~\ref{fig:effPot_tripleTrace0p092}. But since the triple-trace deformation is marginal this is still a CFT, and thus has no phase transitions as a function of temperature.

\paragraph{Single- and triple-trace deformation ($f=0$ and $\Lambda,g\neq 0$)}\label{sec:singleTriple}

\begin{figure}
 \centering
 \includegraphics[scale=0.85,keepaspectratio=true]{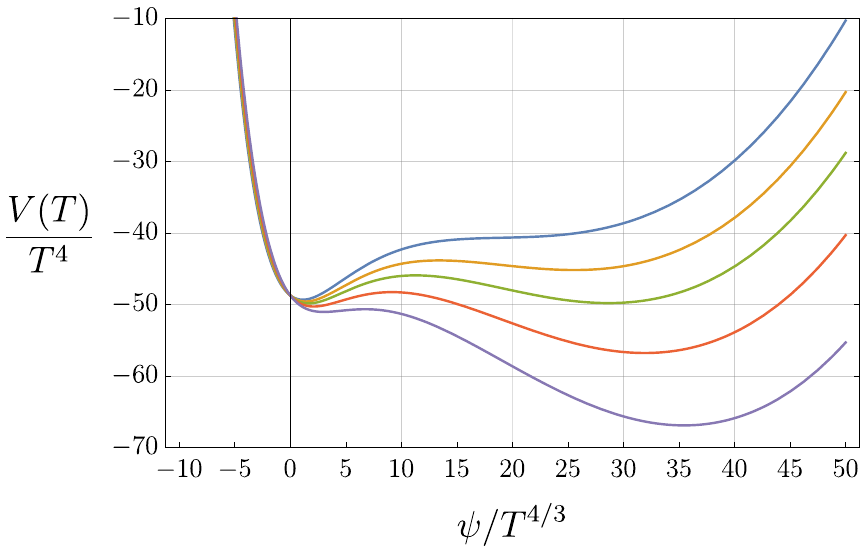}
 \caption{The effective potential for the theory with a single- and triple-trace deformation, for a few different values of $\tilde T$ around the phase transition (at $\tilde T_c\approx 0.844$).The temperature values are $\tilde T\approx$ 0.934, 0.881, 0.844, 0.802, and 0.757 from the top down.}
 \label{fig:effPotTuningSingleTrace}
\end{figure}

Within the interval (\ref{eq:gInt}), the now non-convex potential displays a first order phase transition. The critical temperature depends on $g$; for $g=0.276$, it is $\tilde T\approx 0.844$. With this value of $g$, we also display the potential for a few different $\tilde T$ around the transition in Fig.~\ref{fig:effPotTuningSingleTrace}.

\paragraph{Double- and triple-trace deformation ($\Lambda=0$ and $f,g\neq 0$)}\label{sec:doubleTriple}

\begin{figure}
 \centering
 \includegraphics[scale=0.85,keepaspectratio=true]{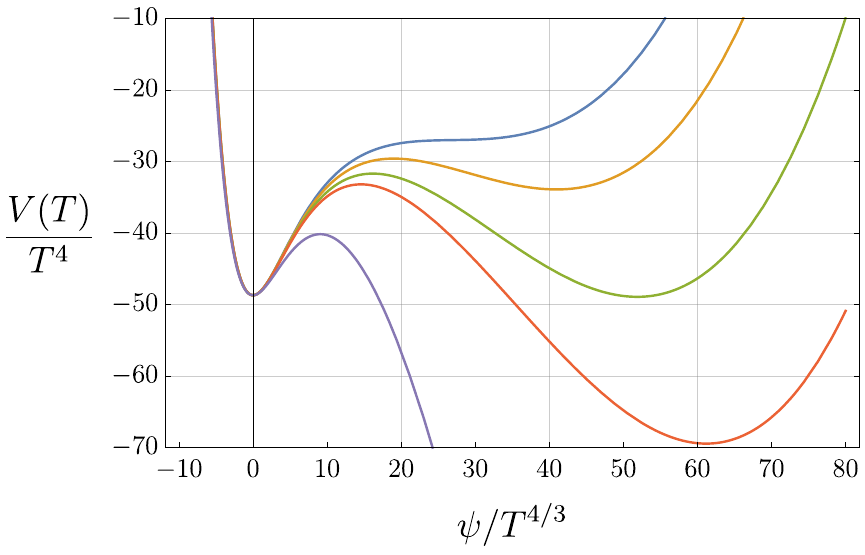}
 \caption{The effective potential for the theory with a double- and triple-trace deformation, for a few different values of $\tilde T$ around the phase transition (at $\tilde T_c\approx 6.72$). The temperature values are $\tilde T\approx$ 8.926, 7.768, 6.724, 6.006, and 3.344 from the top down.}
 \label{fig:effPotTuningDoubleTrace}
\end{figure}

With $g=0$, a negative double trace deformation induces a second order phase transition. Any $0<g<\gamma_3$, however, instead leads to a first order transition. This can be seen by considering the undeformed potential; for small $\psi$, it can be expanded as in (\ref{eq:smallPsiExpansion}); adding the quadratic and cubic contributions from the double- and triple-trace deformations modifies this to be
\begin{equation}
 V(\psi)=V_0 + \frac{1}{2}(V_2 + f) \psi^2 + \frac{g}{3} \psi^3 + \OO(\psi^4) \ .
\end{equation}
Assuming $f>-V_2$, {\emph{i.e.}}, above the temperature $\tilde T$ where the second order phase transition would set in, this cubic potential has a minimum at $\psi=0$, a maximum at
\begin{equation}
 \psi=-\frac{V_2+f}{g} \ ,
\end{equation}
and then dips below $V(0)=V_0$ at
\begin{equation}
 \psi = -\frac{3}{2}\frac{V_2+f}{g} \ .
\end{equation}
If the small-$\psi$ expansion is valid out to this point, this shows that the minimum at $\psi=0$ has become metastable leading to a first order phase transition. But for any given $g\neq 0$, the small-$\psi$ expansion will in fact be valid as long as $f$ is close enough to $-V_2$. This guarantees that as we lower the temperature $\tilde T$ we will always encounter a first order transition \emph{before} the quadratic term goes negative and causes a second order transition. For $g=0.276$, Fig.~\ref{fig:effPotTuningDoubleTrace} shows the effective potential for a few different temperatures around the critical value of $\tilde T\approx 6.72$.

\paragraph{Single-, double- and triple-trace deformation}

In the general case with all couplings non-zero, it is convenient to define a dimensionless coupling
\begin{equation}
 \Lambda_f \equiv \frac{\Lambda}{f^2}
\end{equation}
in addition to the already dimensionless $g$. We then fix the theory by fixing $\Lambda_f$ and $g$, and tune $\tilde T$ to look for a phase transition. The case $\Lambda_f\rightarrow -\infty$ corresponds to the case listed in \ref{sec:singleTriple} with a first order transition within a narrow region of $g$. The case $\Lambda_f=0$ corresponds to the case listed in \ref{sec:doubleTriple} with a first order transition for any $g$, except for $g=0$ which gives a second order transition. The full two-dimensional space of couplings interpolates between these cases. Within some extended region the theory will have a first order phase transition.

\begin{figure}
 \centering
 \includegraphics[scale=1]{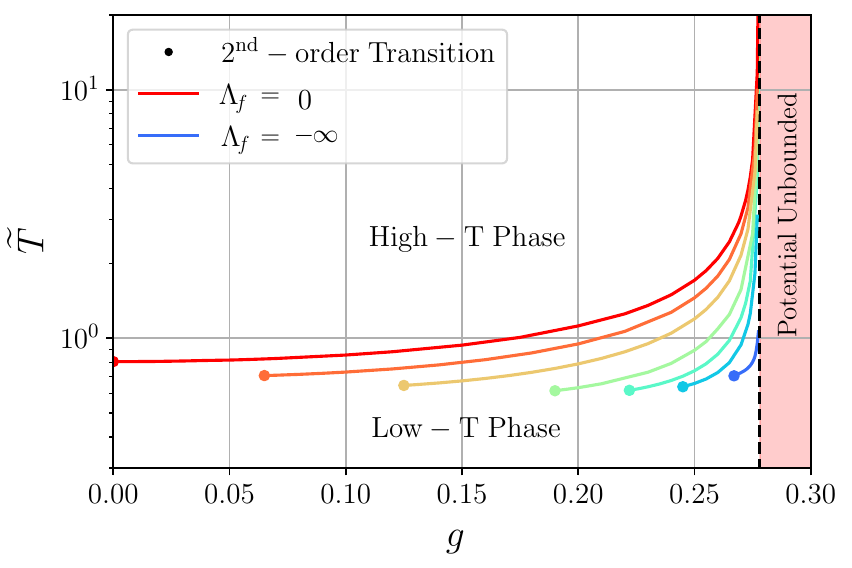}
 \caption{Phase diagram of the dual field theory as a function of $\tilde T$ and $g$, for various values of $\Lambda_f$ ranging from 0 to $-\infty$ and with $N=1$. The curves take values of $\Lambda_f$ = $0$, $-\frac49$, $-1$, $-4$, $-16$, $-100$, and $-\infty$ (top-down).}
 \label{fig:phaseDiagram}
\end{figure}

Fig.~\ref{fig:phaseDiagram} shows a phase diagram of the theory with fixed $\Lambda_f$ as a function of $\tilde T$ and $g$. Each of the coloured curves corresponds to a line of first order transitions for a certain fixed value of $\Lambda_f$; each of these lines terminate in a second order critical point. Note that while we use the label ``high-T phase'' and ``low-T phase'', these are not distinct phases in the sense that one can move from one to the other without any sharp transitions by tuning $g$. The exception is $\Lambda_f=0$, where the first order line extends over the whole allowed range of $g$, except for $g=0$ which corresponds to a second order phase transition.

\subsection{Kinetic term}

Having discussed the effective potential and the resulting phase structure, we move on to the kinetic term, characterised by the function $Z(\psi)$ in (\ref{eq:derivativeExpansionN2}). Note first of all that the non-zero temperature breaks Lorentz invariance, so time and spatial derivatives will not appear on equal footing. We are mainly interested in studying \emph{static} configurations, and so restrict to finding the coefficient of the spatial derivatives; this is what we mean by $Z(\psi)$ in the following.

As discussed in Sec.~\ref{sec:effActionHolography}, we will determine this function by computing the two-point correlator as a function of the expectation value in a low-momentum expansion (at zero frequency). Holographically, this is done by a linearised fluctuation analysis around a given, uniform background solution. We take advantage of the translational symmetry by writing the fluctuations as plane waves, and use the rotational symmetry to align the momentum in the $x$-direction. Making the common gauge choice $H_{M r}=0$, for the metric fluctuation $H_{MN}$ we then have the Ansatz
\bea
ds^{2} & = & -e^{-2\chi(r)}h(r)\left(1+e^{ikx}H_{tt}(r)\right)dt^{2} +\frac{{dr^{2}}}{h(r)}+r^{2}\left(1+e^{ikx}H_{xx}(r)\right)dx^{2} \nonumber \\
      & & + r^{2}\left(1+e^{ikx}H_\perp(r)\right)\left(dy^2+dz^2\right) +2r^{2}e^{ikx}H_{tx}(r)dt\,dx  \\
\phi  & = &  \phi(r)+e^{ikx}\varphi(r) \ .
\eea
Here we have also used the fact that even in the presence of the fluctuations there is an $SO(2)$ rotational symmetry, which restricts which metric components the scalar mode can couple to.

Plugging in this Ansatz into the equations of motion and expanding to linear order, one quickly notes that $H_{tx}$ decouples from the other modes. With a bit more work, $H_{xx}$ can be eliminated, leaving us with three coupled ordinary differential equations (ODEs), first order in derivatives of $H_{tt}$ and second order in derivatives of $H_\perp$ and $\varphi$.

Next, one can show that the following linear combinations of modes are invariant under residual gauge transformations, 
\begin{align}\label{eq:gaugeInvModes}
 Z_\phi(r) &= \varphi(r) - \frac{r}{4}\phi'(r) H_\perp(r) \\
 Z_H(r) &= -e^{-2\chi(r)}h(r)H_{tt}(r) - \frac{r}{4}e^{-2\chi(r)}\left[h'(r)-2h(r)\chi'(r)\right] H_\perp(r) \ ,
\end{align}
and that they satisfy a set of two coupled second order linear ODEs, which we relegate to Appendix~\ref{app:fluctuationEquations} due to their unwieldiness. Near the AdS boundary, these two modes decouple, and the solution asymptotes to
\begin{equation}
 \begin{split}
 Z_\phi(r) &= \frac{Z_\phi^-}{r^{4/3}} + \frac{Z_\phi^+}{r^{8/3}} + \ldots \\
 Z_H(r) &= Z_H^+ r^2 + \frac{Z_H^-}{r^2} + \ldots \label{eq:fluctuationFalloff} \ .
 \end{split}
\end{equation}
From the holographic renormalisation analysis in Appendix~\ref{app:holoRenorm} we see that this small perturbation on the gravity side corresponds to perturbing the source (single-trace coupling) of the dual scalar operator by
\begin{equation}
 \delta \Lambda = Z_\phi^+ + \frac{4}{3} W''\left(\psi\right)Z_\phi^-  \ ,
\end{equation}
where we pick
\begin{equation}
 W(\psi) = \frac{f}{2}\psi^2 + \frac{g}{3}\psi^3 \qquad \text{with} \qquad \psi=-\frac{4}{3}\phi_- \ .
\end{equation}
This perturbation then gives rise to a small change in the scalar expectation value
\begin{equation}
 \delta \langle \Psi \rangle = -\frac{4}{3}Z_\phi^- \ .
\end{equation}
A generic solution to the linearised equations will source not only the scalar mode $Z_\phi$ but also the operator dual to $Z_H$. To compute the scalar two point function, we must then impose the boundary condition $Z_H^+=0$. The two-point function in momentum space is simply the ratio
\begin{equation}
 \langle\Psi\Psi\rangle = \frac{\delta \langle \Psi \rangle}{\delta \Lambda} \ .
\end{equation}

Since we are interested in the low momentum limit of the two-point function, we expand the gauge invariant modes as
\begin{equation}
 Z_i(r) = Z_i^{(0)}(r) + k^2 Z_i^{(2)}(r) + \ldots \qquad \text{with} \quad i\in\{\phi,H\} \ ,
\end{equation}
plug this into the fluctuation equations, and solve order by order in $k^2$. Similarly expanding the coefficients in (\ref{eq:fluctuationFalloff}) as
\begin{equation}
 Z_i^\pm = Z_i^{\pm(0)} + k^2 Z_i^{\pm(2)} + \ldots
\end{equation}
we are able to write the two-point function as
\begin{equation}
 \langle\Psi\Psi\rangle = -\frac{4}{3}\frac{Z_\phi^{-(0)}}{Z_\phi^{+(0)} + \frac{4}{3}W''(\psi)Z_\phi^{-(0)}} - \frac{4}{3}\frac{Z_\phi^{-(2)}Z_\phi^{+(0)}-Z_\phi^{-(0)}Z_\phi^{+(2)}}{\left(Z_\phi^{+(0)} + \frac{4}{3}W''(\psi)Z_\phi^{-(0)}\right)^2}k^2 + \ldots \ .
\end{equation}
Now we use the fact that the part of the effective action quadratic in $\phi$ --- {\emph{i.e.}}, $\Gamma_2$ in the notation of Sec.~\ref{sec:effActionHolography} --- equals the inverse of this two point function. This can be used to derive the following expressions
\begin{equation}\label{eq:Gamma2}
 \Gamma_2 = -\left( \frac{3}{4}\frac{Z_\phi^{+(0)}}{Z_\phi^{-(0)}} + W''(\psi) \right) + \frac{3}{4}\frac{Z_\phi^{-(2)}Z_\phi^{+(0)}-Z_\phi^{-(0)}Z_\phi^{+(2)}}{\left(Z_\phi^{-(0)}\right)^2}k^2 + \ldots \ .
\end{equation}
The first term, of order $k^0$, should just equal the second derivative of the effective potential. We can compare this with the results from the previous subsection to check our numerics; doing so we find excellent agreement. The second term, at order $k^2$, is what we are really interested in as it determines $Z(\psi)$:
\begin{equation}
\label{eq:Z}
 Z(\psi) = \frac{3}{4}\frac{Z_\phi^{-(2)}Z_\phi^{+(0)}-Z_\phi^{-(0)}Z_\phi^{+(2)}}{\left(Z_\phi^{-(0)}\right)^2} \ .
\end{equation}
Note that each $Z_\phi^{\pm(i)}$ here can be regarded as a function of $\psi$, obtained by solving the fluctuation equations in the gravitational background with $\psi=-\frac{4}{3}\phi_-$. Importantly, we note that $Z(\psi)$ is independent of the multi-trace deformations specified by $W(\psi)$, meaning the kinetic term will be the same for the entire class of theories we study. The resulting $Z(\psi)$ is shown in solid red in Fig.~\ref{fig:effPot_undeformed}. It is an even function, admitting an expansion similar to \ref{eq:smallPsiExpansion} for small $\psi$ (large temperatures). At large $\psi$ (small temperatures) it goes as $\psi^{-1/2}$ (see the dashed black curve), which is required by scale invariance to make the kinetic term have dimension four.

\subsection{Higher-derivative terms}\label{subsec:k4}

It is straightforward to continue the work of the previous subsection and compute the two-point function up to higher order in $k^2$. By the same reasoning as above, this should provide information about higher-derivative terms in the effective action. A complication arises though, since starting at four derivatives, the number of independent terms grows rapidly. Some of these terms involve several external momenta, and require information from higher-order correlation functions to determine. Computing these is beyond the scope of this paper. We have, however, computed the two point function up to order $k^4$, which lets us extract the four-derivative term $\nabla^2\psi\nabla^2\psi$. This allows us to verify that this term is negligible in the context of bubble nucleation --- discussed in the next section --- thus providing evidence that the small derivative expansion is applicable there.

\section{Bubble nucleation and $N$ dependence}\label{sec:bubbles}

One important motivation for going after the effective action is to understand the phase structure and phase transitions of a field theory. First order phase transitions proceed through bubble nucleation, which in turn is controlled by the effective action previously found. At an arbitrary temperature, the resulting equations of motion should be solved in Euclidean space with the appropriate periodicity imposed in the time-direction, leading in general to a \emph{partial} differential equation. Typically this is simplified into an ordinary differential equation by focussing on the high and low temperature limits, 
leading to two actions: a zero temperature action with O(4) symmetry arising from purely quantum fluctuation effects causing bubble nucleation \cite{Coleman:1977py,Callan:1977pt} and a non-zero temperature, O(3) symmetric action arising from both thermal and quantum fluctuations \cite{Linde:1981zj}.

Note that while an O(3) symmetric typically exists for any temperature, the O(4) solution only exists at zero temperature, where Lorentz symmetry is unbroken. Nonetheless, it is a good approximation to a true solution as long as the length of the thermal circle is large (and thus the temperature small) compared with the bubble radius. Motivated by this, we study both O(3) and O(4) solutions. We assume that the function multiplying the kinetic term in the O(4) case is the same as in the static O(3) case, which is what was computed in the previous section. This is a reasonable assumption since the O(4) solution is expected to matter more at low temperatures, where Lorentz symmetry is nearly restored.

After integrating along the angular direction in Euclidean spacetime, the $O(4)$ symmetric action is
\be
 \Gamma_{O(4)} = 2\pi^2 N^2\int_0^{\infty}d\rho\, \rho^3\left(\frac12Z(\psi)\left(\frac{d\psi}{d\rho}\right)^2+V(\psi)\right) \ ,
\end{equation}
while for the $O(3)$ symmetric action we integrate along the angular spatial directions and the Euclidean time circle
\be
 \Gamma_{O(3)} = \frac{4\pi N^2}{T}\int_0^{\infty}d\rho\, \rho^2\left(\frac12Z(\psi)\left(\frac{d\psi}{d\rho}\right)^2+V(\psi,T)\right) \ .
\end{equation}
Note that in each case $\rho$ takes a different meaning, in the $O(4)$ symmetric configuration it is the radial direction in full Euclidean four-dimensional spacetime, while in the $O(3)$ symmetric configuration $\rho$ is the radial direction along the three-dimensional space. As earlier, we pull our a factor of $N^2$, all other quantities then being of order $N^0$.

\begin{figure}[h]
  \center
      \includegraphics[width=0.6\textwidth]{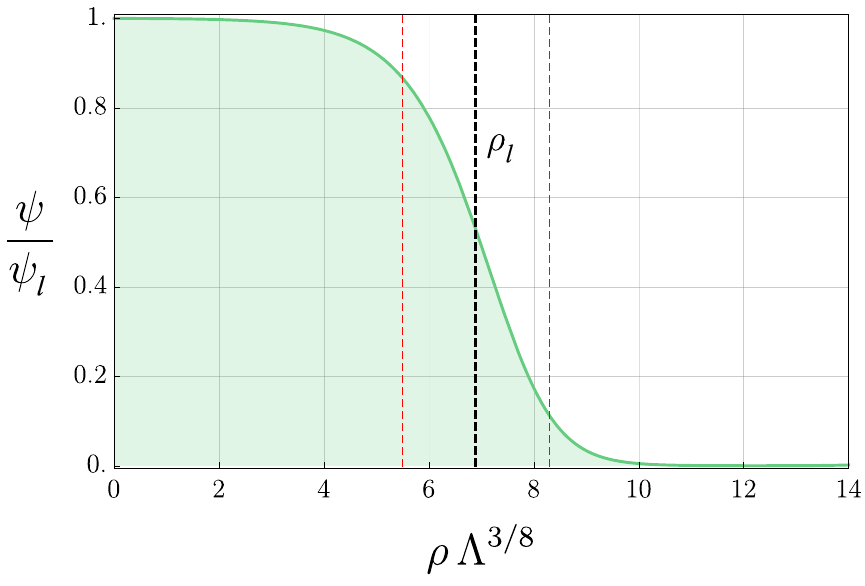}
  \caption{Graph of an $O(3)$ bubble solution with $N=1$ normalised by the low-$T$ phase value $\psi_l$ against the radius. The thick dashed vertical black line in the middle denotes the radius of the bubble $\rho_b$, and the vertical dashed red lines encompass the bubble thickness. The parameter values are $g=0.27$, $\Lambda_f = 0$.}
  \label{fig:bubsol}
\end{figure}

To evaluate these integrals we first need to find the field profiles $\psi(\rho)$, by solving the equations of motion derived from the effective action. For the $O(4)$ symmetric bubble, the equation of motion is
\be
\frac{d^2\psi}{d\rho^2}+\frac{3}{\rho}\frac{d\psi}{d\rho}+\frac12\frac{\partial_\psi Z(\psi)}{Z(\psi)}\left(\frac{d\psi}{d\rho}\right)^2-\frac{\partial_\psi V(\psi)}{Z(\psi)} = 0 \ ,\ee
while for the $O(3)$ symmetric bubble is
\be\label{O3EOM}
\frac{d^2\psi}{d\rho^2}+\frac{2}{\rho}\frac{d\psi}{d\rho}+\frac12\frac{\partial_\psi Z(\psi)}{Z(\psi)}\left(\frac{d\psi}{d\rho}\right)^2-\frac{\partial_\psi V(\psi)}{Z(\psi)} = 0 \ .
\ee
These are solved with the well-known ``shooting method" with boundary conditions $\psi(\infty)=\psi'(\infty)=0$ where the initial minimum is shifted to always appear at $V = 0$. An example of a bubble profile is shown in Fig. \ref{fig:bubsol} which solves the O(3) equation of motion \eqref{O3EOM}. 

How steeply the profile transitions from one phase to another details the ``thickness" of the bubble wall. Whether the wall is in the thin, thick, or intermediate regime will determine how important quantities will change when considering the number of colours $N$ of the theory, as shown shortly.

\begin{figure}[ht!]
  \center
      \includegraphics[width=0.65\textwidth]{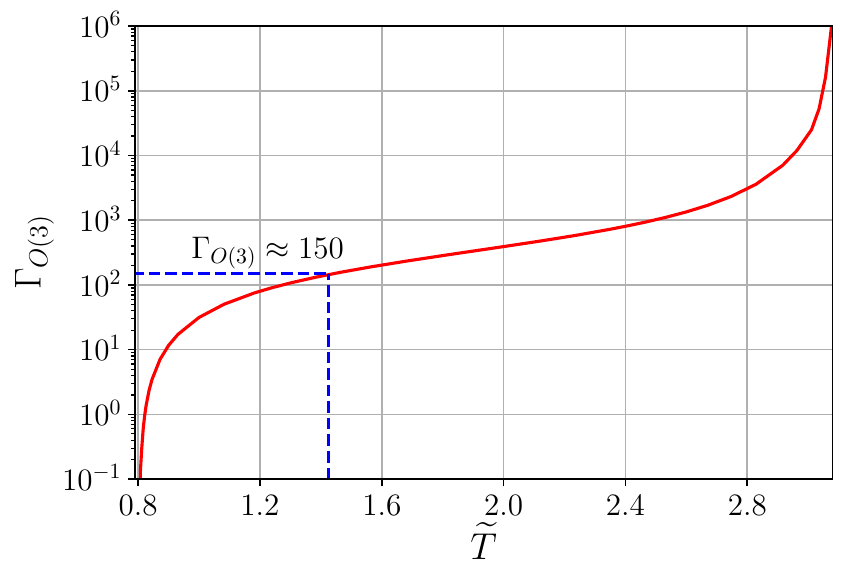}
      \caption{Graph of the $O(3)$ bubble action against the temperature $\widetilde{T}$ for triple-trace coupling $g=0.27$, $\Lambda_f = 0$, and with $N=1$.}
      \label{fig:O3Bub}
\end{figure}

In our holographic model, a complete picture of how the bubble action depends upon the temperature is built up by selecting a particular value of the triple trace coupling $g$ and coupling ratio $\Lambda_f$, then varying the temperature $\widetilde{T}$. There is therefore one graph of the action for each parameter set $g, \Lambda_f$, all of which appear similar in shape to Fig.~\ref{fig:O3Bub}.

\begin{figure}[ht!]
 \centering
  \includegraphics[width=0.8\textwidth]{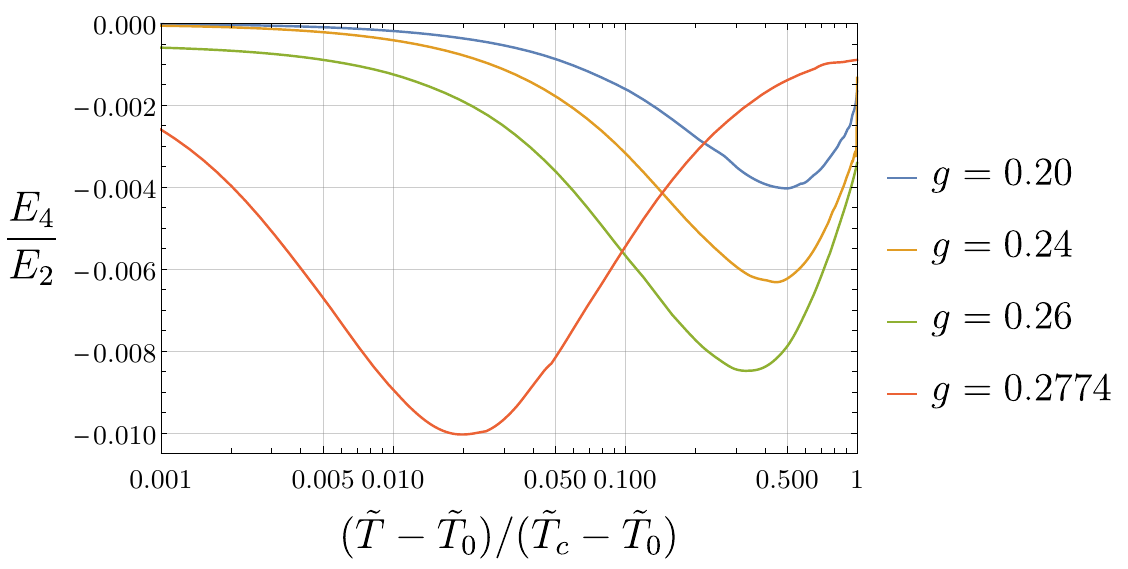}
  \caption{Graph of the ratio of $k^4$ and $k^2$ energy contributions in the kinetic coefficient $Z(\phi)$ against the temperature $\widetilde{T}$ for triple-trace couplings $g = 0.20$, 0.24, 0.26, and 0.2774 (top-down on the left), with $\Lambda_f = 0$. All curves are for value $N=1$.}
  \label{fig:k4vsk2}
\end{figure}

Another aspect which must be taken into consideration is the fact that we have used a small derivative (or low momentum) expansion for the effective action, truncated at two derivatives. As it is not immediately clear whether the higher order terms will be negligible, we explored what consequence including the term $\partial^2\psi\partial^2\psi$ has on the bubble action. As discussed in subsection \ref{subsec:k4}, this term in the effective action can be obtained by computing the scalar two-point function to order $k^4$.  The importance of the effect can be judged by the size of the ratio $E_4/E_2$, where $E_2$ is the contribution to the energy from including the $k^2$ term and $E_4$ is the contribution to the energy from including the $k^4$ term, shown in Fig.~\ref{fig:k4vsk2}.  This is plotted against the scaled temperature defined as $(\tilde{T}-\tilde{T}_0)/(\tilde{T}_c-\tilde{T}_0)$, where $\tilde{T}_c$ is the dimensionless critical temperature, and $\tilde{T}_0$ is the dimensionless lower critical temperature at which the metastable minimum disappears.

As can be seen in this figure $E_4$ is a negative contribution with magnitude of less than $1\%$ of the $k^2$ contribution, and this remains valid in general. We therefore determine that the higher derivative terms can be safely ignored, at least for the correction quadratic in the fields, and that the $k^2$ term will probably give a very good approximation to the true kinetic function $Z(\psi)$.

\begin{figure}[ht!]
  \center
      \includegraphics[width=0.8\textwidth]{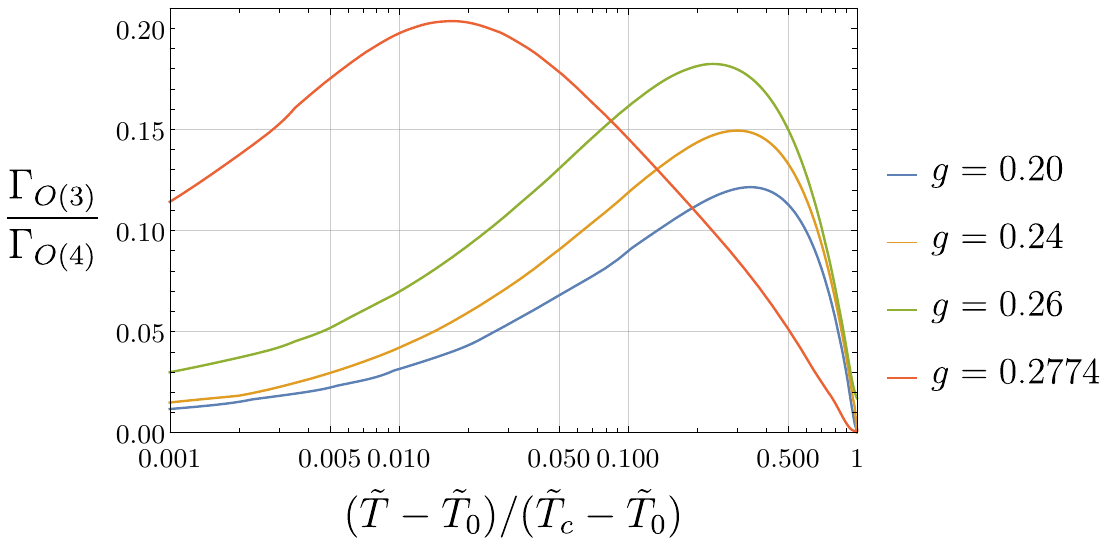}
      \caption{Graph of the ratio of $O(3)$ bubble action over $O(4)$ bubble action against the temperature $T$ in units of the source $\Lambda$ for triple-trace couplings $g = 0.20$, 0.24, 0.26, and 0.2774 (bottom-up on the left), with $\Lambda_f = 0$. All curves are for value $N=1$.}
      \label{fig:O3vO4}
\end{figure}

The quantities important in the companion paper \cite{shortcompanion} all rely upon the characteristics of the action curve demonstrated by Fig. \ref{fig:O3Bub}, and in particular which action fulfils the criterion
\be
\Gamma(T) = \textrm{min}[\Gamma_{O(3)}(T),\Gamma_{O(4)}(T)] \ ,
\ee
as this could drastically change the temperature at which the bubble nucleates and which region of thickness the bubble is in. We therefore also perform the check of calculating the $O(3)$ and $O(4)$ bubble actions for each temperature value and comparing the sizes, with one illustration of this seen in Fig.~\ref{fig:O3vO4}. The action obtained from the $O(3)$ bubble is consistently significantly lower than from the $O(4)$ bubble, and thus will dominate the calculation of the subsequent quantities.

\subsection{Phase transitions in the early universe and large-$N$ scaling}

Much recent effort --- including that of our companion paper \cite{shortcompanion} and previous paper \cite{Ares:2020lbt} --- has gone into modelling phase transitions in the early universe using holography, with the hope of finding models that lead to observable gravitational wave signals. In this context, it is important to understand the impact of the large-$N$ limit that holography usually involves. As mentioned in Sec.~\ref{sec:holography}, the (effective) action obtained from holography scales as $L^3/\kappa_5^2 \sim N^2$, meaning that in the strict large-$N$ limit bubble nucleation will not occur as the bubble action becomes infinite. In practice we are of course interested in $N$ values which are finite, while still being large enough to ignore finite-$N$ corrections.

\begin{figure*}[ht!]
  \center
      \includegraphics[width=0.4768\textwidth]{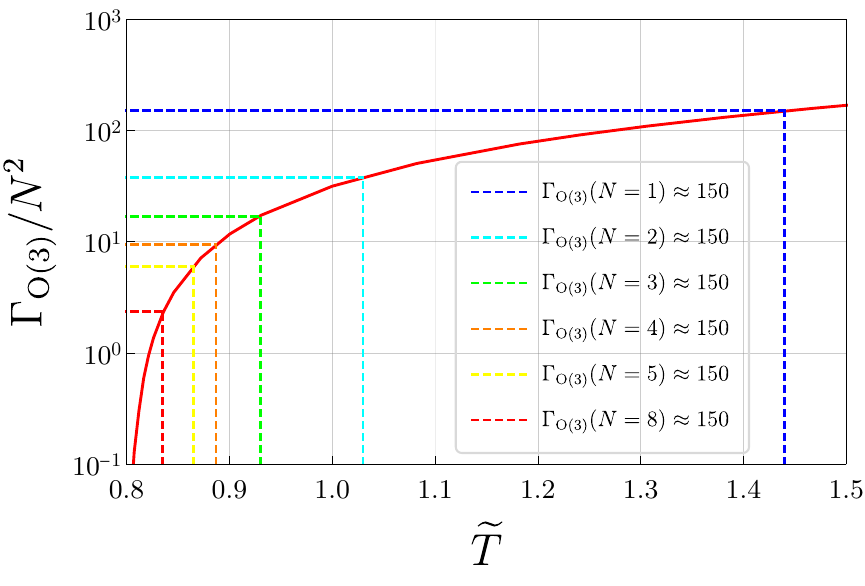}
      $\;\;\;\;\;$ 
      \includegraphics[width=0.4768\textwidth]{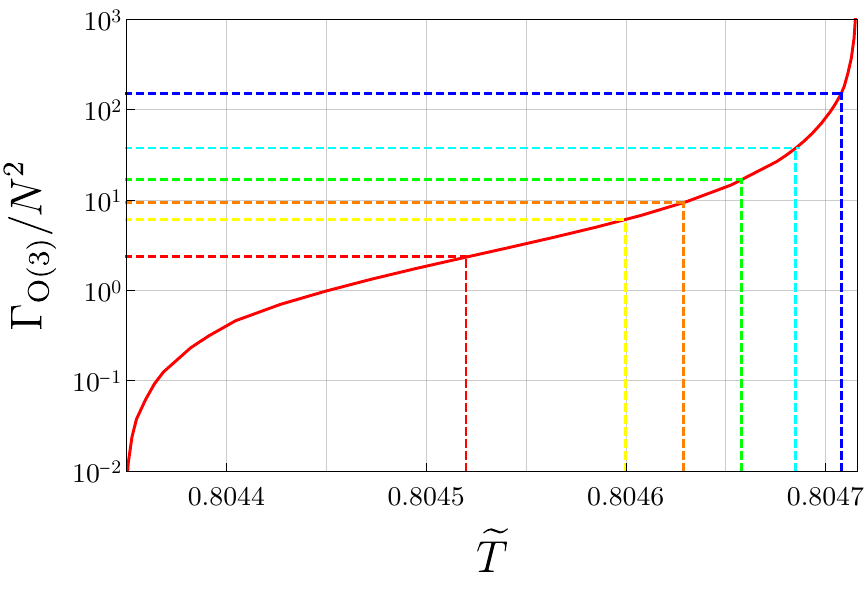}
      \caption{The solid red curves show a section of the graph of the $O(3)$ bubble action against the temperature $\widetilde{T}$ for $g=0.27$ and $\Lambda_f = 0$ (left), and the complete graph for $g=0.01$ and $\Lambda_f = 0$ (right). As $N$ is increased, the point where the action equals 150, which sets the nucleation temperature, is pushed to the left. The legend applies for both plots.}
      \label{fig:actionscaled}
\end{figure*}

In cosmological applications, important quantities which are determined from the effective action include the nucleation temperature $T_n$, the transition strength $\alpha$, and the transition rate $\beta/H_n$. The nucleation temperature is defined as the temperature at which bubble nucleation will occur (specifically when the nucleation rate per unit volume drops to one bubble per Hubble volume per Hubble time). Through energy considerations (see \cite{shortcompanion}) this is found to always occur at $\Gamma\approx150$, and so changing the scale of the action by changing $N$ will invariably alter the temperature at which bubbles are nucleated. We demonstrate this in Fig.~\ref{fig:actionscaled} for various values of $N$ up to $N=8$, which is the value we take to produce results in the companion paper. ($N\approx 1$ is of course outside the plausible range of the large-$N$ expansion; here we are mainly interested in visualising the impact of changing $N$ on the cosmological parameters.)

\begin{figure*}[ht!]
  \center
      \includegraphics[width=0.461\textwidth]{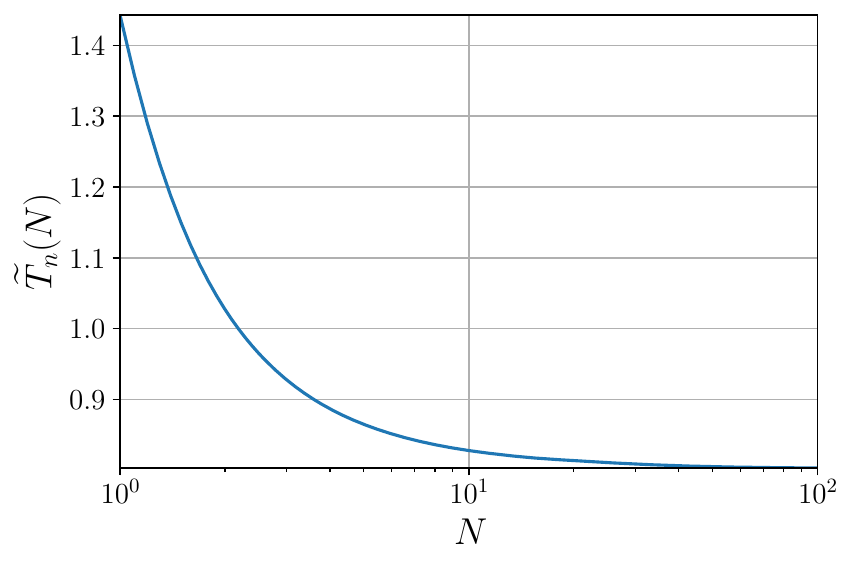}
      $\;\;\;\;\;$ 
      \includegraphics[width=0.491\textwidth]{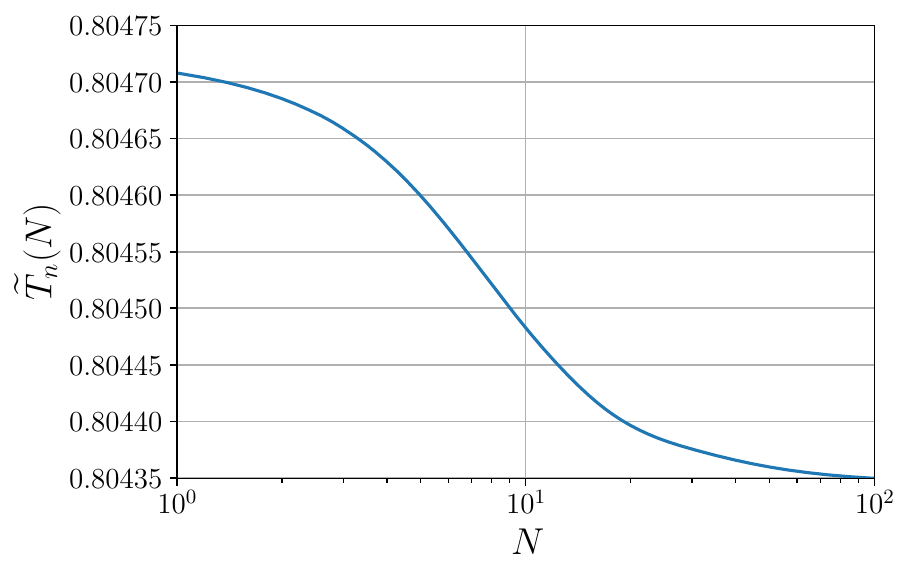}
      \caption{Graph of the nucleation temperature $\widetilde{T}_n$ against $N$ for triple-trace coupling $g=0.27$ and $\Lambda_f = 0$ (left), and $g=0.01$ and $\Lambda_f = 0$ (right) showing the effect of $N$ scaling.}
      \label{fig:Tnscaled}
\end{figure*}

It is very straightforward to numerically invert the action curve to find out how the nucleation temperature depends upon $N$. In Fig. \ref{fig:Tnscaled} we show the nucleation temperature dependence upon $N$ for the same two sets of parameters as in Fig. \ref{fig:actionscaled}, demonstrating the change in shape as the $N$ scaling progresses from near the thin-wall limit to the thick-wall limit. We note in particular that as $N\to\infty$, the nucleation temperature always approaches the lower critical temperature $T_0$, where the barrier between the vacua in the effective potential vanishes and the bubble action goes to zero.

The transition rate $\beta/H_n$ can be expressed as
\be\label{B_H}
\frac{\beta}{H_n} = T\frac{d}{dT}\Gamma(\psi,T)\Big|_{T_n} \ ,
\ee
where $H_n$ is the Hubble parameter evaluated at the nucleation temperature.
As the transition rate actively involves a derivative with respect to $\Gamma$, the $N^2$ scaling of the holographic action has a substantial effect. For sufficiently large $N$, we have as already mentioned $T_n\approx T_0$ placing us in the thick wall limit. We observe that the temperature dependence of the action in our model is well fitted by a power law near $T_0$,
\be
 \Gamma\sim N^2(T-T_0)^x
\ee
with $x>0$. Using this and the fact that $T_n$ occurs when $\Gamma\approx 150$, we get
\be
 N^2(T_n-T_0)^x \sim 150 \ .
\ee
Then, using this in Eq.~(\ref{B_H}), we find
\be
 \frac{\beta}{H_n} \sim T_n N^2 x\, (T_n-T_0)^{x-1} \sim T_n N^2 x (150 N^{-2})^{\frac{x-1}{x}} \sim 150^{\frac{x-1}{x}}T_n x N^{2/x} \ , 
\ee
implying that $\beta/H_n \sim N^{2/x}$. Thus, for sufficiently large $N$, the transition rate diverges rendering the gravitational wave signal unobservable.

However, for more moderate values of $N$ it is possible to instead sit close to the thin wall limit, $T_n\approx T_c$. The quadratic divergence in this limit, $\Gamma\sim N^2(T_c-T)^{-2}$, then leads in a similar way to $\beta/H_n \sim N^{-1}$, decreasing with $N$ in accordance with lattice results for the surface tension \cite{Lucini:2003zr}.

\begin{figure*}[ht!]
  \center
      \includegraphics[width=0.4768\textwidth]{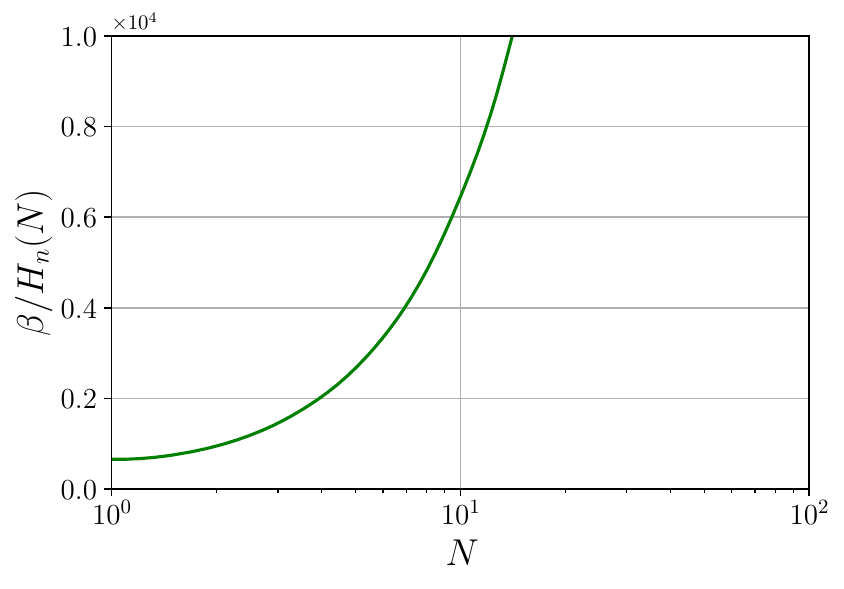}
      $\;\;\;\;\;$ 
      \includegraphics[width=0.4768\textwidth]{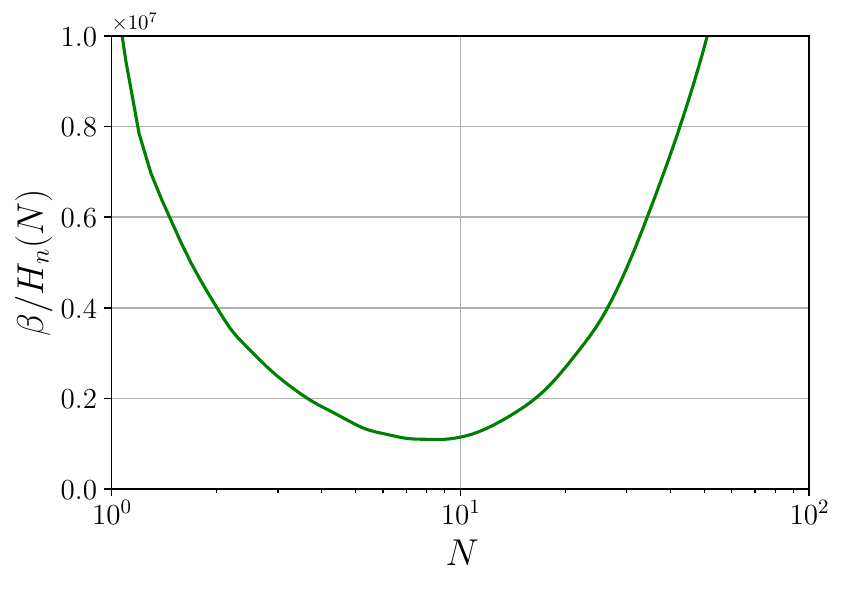}
      \caption{Graph of the transition rate $\beta/H_n$ against $N$ for triple-trace coupling $g=0.27$ and $\Lambda_f = 0$ (left), and $g=0.01$ and $\Lambda_f = 0$ (right) showing the effect of $N$ scaling.}
      \label{fig:betascaled}
\end{figure*}

The above analytic results for the $N$ scaling in the thin and thick wall limit can be confirmed numerically, as shown in Fig. \ref{fig:betascaled}. On the left, we see that as the action at $\Gamma\approx150$ for $N=1$  begins close to the lower critical temperature $T_0$ for $g=0.27$, it is dominated by the thick-wall $N$ scaling result, always increasing the transition rate with $N$. For $g=0.01$ on the right, however, we see that the action at $\Gamma\approx150$ for $N=1$ begins in the thin-wall regime close to $T_c$, with $\beta/H_n$ decreasing as $N$ increases slightly, then as $N$ increases further it transitions through the intermediate and then into the thick-wall regime where it again ends up dominating and increasing the transition rate. In this case there is an ``optimal" $N\approx 8$ which minimises the transition rate (leading to easier to observe gravitational wave signals), while still potentially being large enough to avoid significant finite-$N$ corrections.

\subsection{Domain wall solutions}

As a special case of the study of bubble nucleation, we can also consider the limit $T\to T_c$, often referred to as the thin-wall limit since the bubble wall here becomes small compared to the overall size of the bubble. Precisely at $T=T_c$ there is no bubble nucleation (as the bubble action diverges) but one can instead have coexistence of the two phases, separated by a domain wall with some particular surface tension. Determining the surface tension is interesting both since it can be related to the nucleation rate near $T_c$ and since it determines properties of possible mixed phases. Domain walls in mixed phases have already been studied using dynamical solutions of the Einstein equations \cite{Attems:2017ezz,Janik:2017ykj,Attems:2019yqn,Bellantuono:2019wbn,Li:2020ayr,Bea:2021zsu} and with an effective action phenomenologically derived from a holographic model \cite{Janik:2021jbq}. 
Here, we investigate the domain wall solutions using an effective action which is rigorously derived in a gradient expansion using the rules of holography. 

\begin{figure}[h!]
 \centering
  \includegraphics[width=0.75\textwidth]{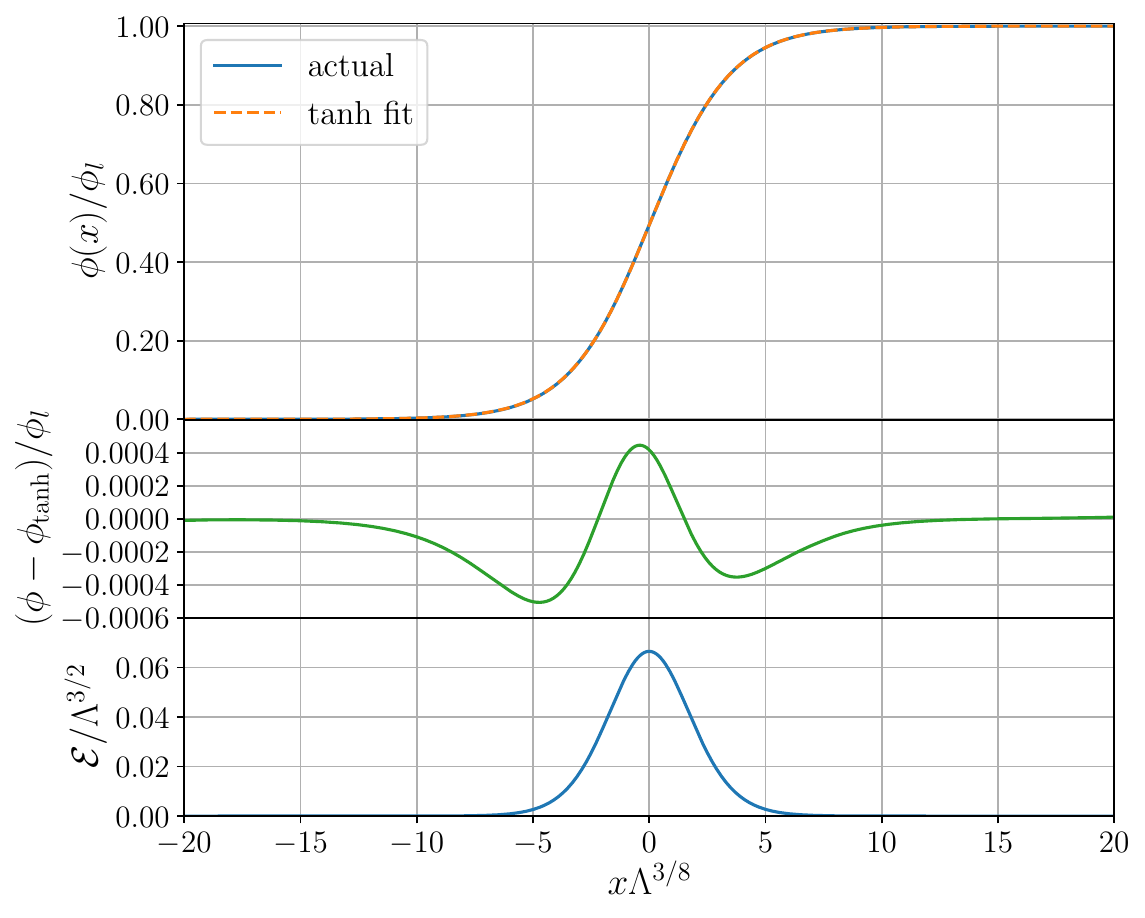}
  \caption{Profile of the difference between the field and its high-$T$ phase minimum $\phi = \psi - \psi_h$ normalised by its maximum value $\phi_l$, the  thin-wall approximation to the profile, the difference between the two, and the energy of the profile for $g = 0.1, \Lambda_f = 0$. All curves are for value $N=1$.}
  \label{fig:tanh}
\end{figure}

The surface tension $\sigma$ is calculated through the formula
\begin{equation}\sigma = \int^{\infty}_{-\infty}dx\left(\frac12 Z(\phi(x))\left(\nabla\phi(x)\right)^2+V(\phi(x))\right) \ ,
\label{eq:sigma}
\end{equation}
which represents the surface tension of a domain wall extending along the $x = 0$ plane, with the domain wall field profile here being related to the bubble field profile through $\phi = \psi-\psi_h$ at $T_c$, where $\psi_h$ is the value of the field at the high-$T$ phase minimum. This integral can be performed by finding the profile of the field $\phi(x)$ numerically through the equation of motion
\begin{equation}
\frac{d^2\phi}{dx^2}+\frac12\frac{\partial_\phi Z(\phi)}{Z(\phi)}\left(\frac{d\phi}{dx}\right)^2-\frac{\partial_\phi V(\phi)}{Z(\phi)} = 0 \ .
\end{equation}

We utilise $\it{Mathematica's}$ $\bf{FindRoot}$ function to first locate the extrema in our effective potential, and then the $\bf{NDSolveValue}$ function with the shooting method. We shoot from the local maximum to each of the minima, providing an initial guess of the derivative of the field at the local maximum and varying it until the solution reaches the minimum and stays there for a sufficient distance. 

It is interesting to compare the resulting solution to a simple analytic approximate expression with Lagrangian $\mathcal{L} = \frac12 \left(\partial_\mu\phi\right)^2-V(\phi)$, which is exact when the potential is an even quartic function. The domain wall solution to this then takes the form of a hyperbolic tangent \cite{Janik:2021jbq},
\begin{equation}
\phi_{\mathrm{tanh}}(x) = \frac{\phi_b}{2}\left(1+\mathrm{tanh}\left(\frac{x}{L_w}+\delta\right)\right) \ .
\end{equation}
Here $\phi_b$ is the value of the field in the broken minimum, $L_w$ is the thickness of the wall, and $\delta$ is a parameter allowing for a shift in the domain wall. Once we have the field profile from the numerical solver, we can use a fitting function such as $\it{Python's}$ $\bf{curve\_fit}$ to find the values of $L_w$ and $\delta$ which best approximate the actual solution.

\begin{figure}[ht!]
 \centering
  \includegraphics[width=0.75\textwidth]{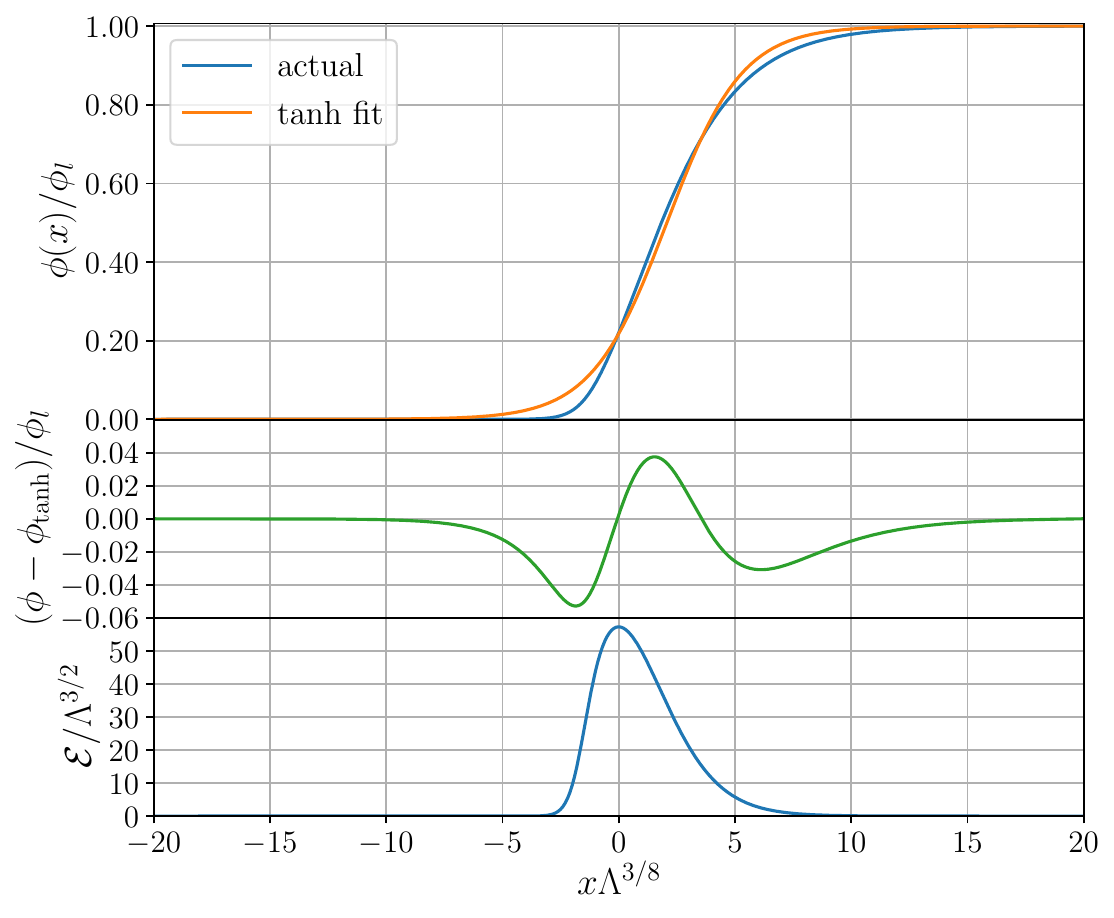}
  \caption{Field profile and thin-wall approximation, difference between the two, and energy of the profile for $g = 0.2774, \Lambda_f = 0$. All curves are for value $N=1$.}
  \label{fig:tanh2}
\end{figure}

\begin{figure*}[ht!]
  \center
      \includegraphics[width=0.485\textwidth]{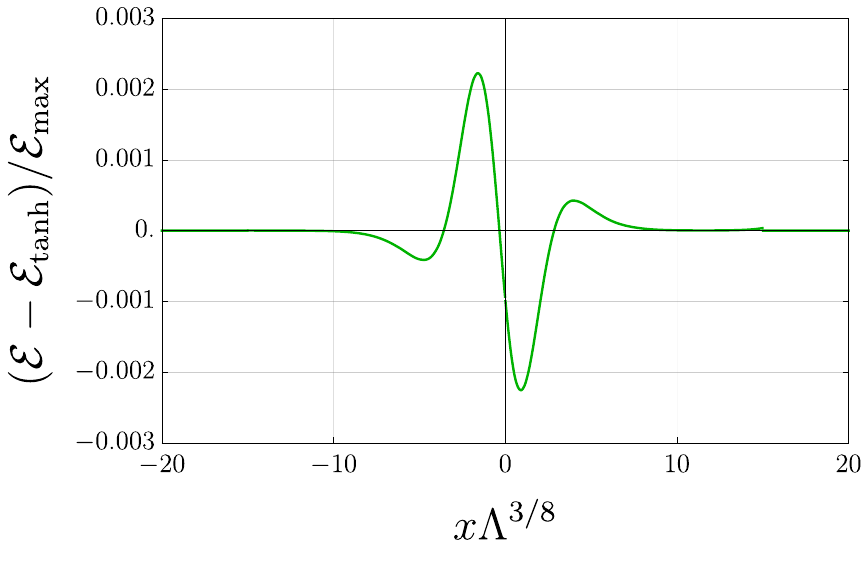}
      $\;\;\;\;\;$ 
      \includegraphics[width=0.471\textwidth]{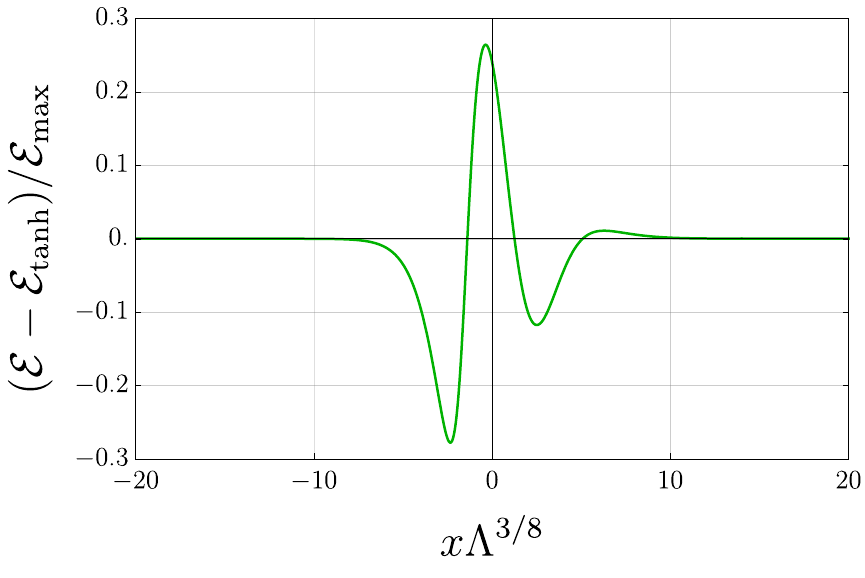}
      \caption{Plots of the difference between the actual energy density solution $\mathcal{E} $and the tanh approximation energy density solution $\mathcal{E}_{\textrm{tanh}}$ scaled by the maximum value of the actual energy density $\mathcal{E}_{\textrm{max}}$ for the cases $g =1, \Lambda_f = 0$ on the left and $g = 0.2774, \Lambda_f = 0$ on the right. Both curves are for value $N=1$.}
      \label{fig:Energydiffs}
\end{figure*}

In Fig. \ref{fig:tanh} we plot the analytic tanh-solution over the actual solution for parameter set $g = 0.1$, $\Lambda_f = 0$, along with the difference between the two solutions in the form of $(\phi(x)-\phi_{\textrm{tanh}}(x))/\phi_l$ and the scaled energy density $\mathcal{E}$ in units of the source $\Lambda$, where the energy density is given through
\begin{equation}
\mathcal{E} = \frac12 Z(\phi(x))\left(\nabla\phi(x)\right)^2+V(\phi(x)) \ .
\end{equation}
The effective potential for this parameter set has minima which are close together, and the curve between the minima is well-approximated by a quartic function, which is also seen in how closely the tanh approximation overlays the actual field profile. As is evident from the middle plot of Fig. \ref{fig:tanh}, the difference between the approximate solution and the actual solution for the scalar field is never greater than $5\times10^{-4}$. The difference in energy density $\mathcal{E}$ between actual and tanh-fit is similarly good, never increasing past $3\times10^{-3}$ (seen in Fig. \ref{fig:Energydiffs}).

In Fig. \ref{fig:tanh2} however for parameter set $g = 0.2774$, $\Lambda_f = 0$, the minima of the effective potential are far away from each other, and the curve between the minima is not close to a simple quartic. Thus the tanh-fit performs significantly worse here; the differences between the actual and approximate solution are about 100 times larger than in the previous case. Another interesting feature for this parameter set in Fig.~\ref{fig:tanh2} is how skewed the energy density is in the bottom plot. For domain walls taking a tanh-like form we see a nicely symmetric energy distribution like in Fig.~\ref{fig:tanh}, which again demonstrates how far the departure is from this approximation. This departure is greatly pronounced for $\cal{E}$, with the difference in results off by tens of percents (close to $30\%$ difference where the effect is most noticeable). This is again about 100 times larger than in the previous case and is also seen in Fig. \ref{fig:Energydiffs}.

\begin{figure}[h!]
 \centering
  \includegraphics[width=0.75\textwidth]{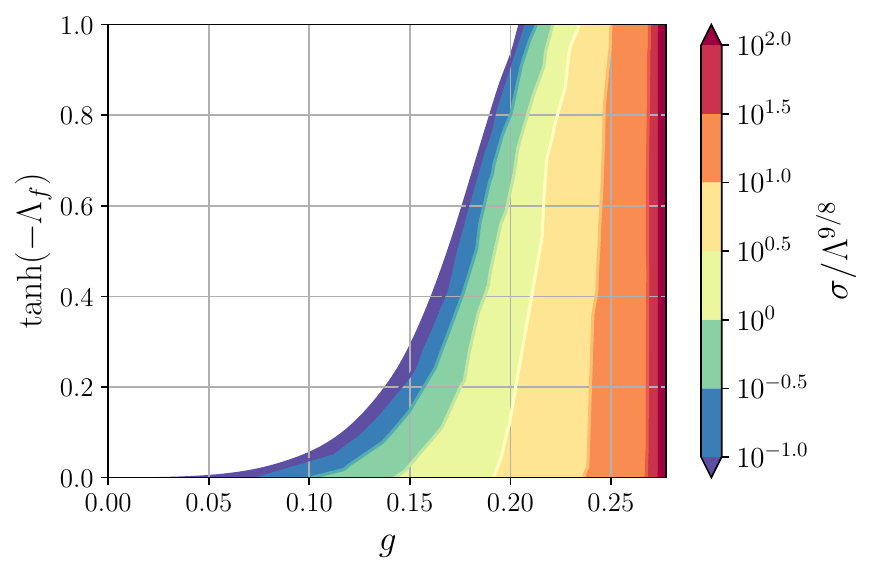}
  \caption{Scan of the domain wall surface tension $\sigma$ in units of the source, with $N=1$.}
  \label{fig:sigma}
\end{figure}

Once we have found how the field behaves between the two minima, we can input it into eq. \ref{eq:sigma} to be able to calculate the surface tension. We produce a scan of this quantity in Fig. \ref{fig:sigma}, with the surface tension in units of the source. Near the left hand boundary where $g$ is small and the minima are close together we see that the surface tension is small, and goes explicitly to zero at the boundary. On the right hand side, however, where $g$ is large and the minima are far away from one another the surface tension reaches values over 100. In this area of parameter sets, the values found effectively drop their dependences on $\Lambda_f$ as the minima are so far apart that the difference is negligible. A note here is that although the surface tension will scale with $N$, it scales uniformly and so will not distort different parts of the parameter space in different ways. 

\section{Discussion and outlook}\label{sec:discussion}

In this paper we discussed the computation of the quantum effective action in a strongly coupled theory using holographic duality. The effective action was computed in a derivative expansion, and we focused on extracting the effective potential and the (non-canonical) two-derivative kinetic term. Higher-derivative terms can in principle be included in a straightforward fashion; however, the number of possible terms grows quickly at higher order, and higher order correlation functions must be computed, requiring considerably more effort. 

Our methods for constructing the effective action are general, applying to any holographic model. It is however interesting to note a possible complication. One of the ways to construct the effective potential was to integrate the source $J$ with respect to the field $\psi$, implying that $J$ should be a single-valued function of $\psi$.  We know that there is a one-parameter family of solutions to the bulk theory, meaning that one can construct a curve in the $(\psi,J)$ plane, but 
we know of no reason why the function $J(\psi)$ is guaranteed to be single-valued, as it happens to be in the simplified model.

As a concrete example of our approach we studied a simple bottom-up gravity theory, with a scalar field whose mass allows it to be identified with a dimension-4/3 operator in the dual field theory. Turning on a temperature and deforming the putative dual CFT by single-, double-, and triple-trace operators, we mapped out a surprisingly rich phase diagram. 

Our main motivation for computing the effective action was to study first order phase transitions mediated by bubble nucleation. Thus we proceeded to find ``bounce'' solutions to the equations of motion obtained from the effective action, and studied their properties. An interesting application of this technology is to early-universe cosmology, where a first-order phase transition can give rise to potentially observable gravitational waves. Our companion paper \cite{shortcompanion} will discuss this in more detail, including the computation of all quasi-equilibrium gravitational wave parameters in our simple toy model.

Bubble nucleation is based on the idea of fluctuations, quantum or thermal, which allow the system to overcome the potential barrier between the false and true vacua. In the gravity dual, such fluctuations are suppressed by the large parameter $L^3/\kappa_5^2$. For a holographic theory to make an {\emph{observable}} prediction in the case of gravitational waves signals (for example), it is thus vital to set this parameter (which is roughly dual to the degrees of freedom, or $N^2$ in a gauge theory) to some finite value. With this in mind, we briefly discussed the $N$-dependence of our results. The main takeaway is that, while indeed the transition rate always diverges for $L^3/\kappa_5^2\sim N^2$ large enough, there can in general exist a range of $N$ where the transition rate is somewhat \emph{suppressed} compared with the naive extrapolation to $N=1$.

As we elaborate on in more detail in the companion paper \cite{shortcompanion}, the simple holographic model studied herein leads, for most of the parameter range, to large transition rates, implying difficult-to-observe gravitational wave signals. It would be interesting to explore more models in this way, and ideally isolate properties of the gravity dual leading to observable signals.

In addition to cosmological applications, the framework outlined here might also find uses in holographic models of nuclear physics at non-zero charge density \cite{Bergman:2007wp,Jokela:2020piw,Kovensky:2020xif}. Here, one would be concerned with, {\emph{e.g.}}, the possible first order transition between the nuclear and chirally restored quark matter phases. If no other scalar condensation would occur in this transition, the order parameter would simply be the charge density, jumping from one non-zero value to another. The corresponding source would then be the chemical potential. However, since chiral symmetry will be restored, a careful implementation of holographic renormalisation is crucial when identifying the source and the expectation values of the dual operator, and therefore also in the computation of the effective action.

The chiral transition is particularly interesting in astrophysical contexts. For example, to address the question if the quark matter phase is realisable in stellar processes, one needs to know the relevant time scales of the phase conversion process. In addition to the pressure difference between the phases, a key ingredient setting the time scale is the surface tension \cite{Mintz:2009ay}, the computation of which we have also discussed here. Indeed, the surface tension is relevant for bubble nucleation of quark matter in supernovae \cite{Sagert:2008ka,Logoteta:2012ms}, neutron star mergers \cite{Most:2018hfd,Bauswein:2018bma,Chesler:2019osn}, and for a possible quark-hadron mixed phase in the interior of quiescent neutron stars \cite{Glendenning:1992vb,Heiselberg:1992dx}.

Ideally, as with the equation of state, the surface tension should be calculated from the underlying fundamental theory, QCD. The density regimes, where two available perturbative approaches, chiral effective theory and perturbative QCD, are valid are far apart such that at a possible first-order transition at least one of them, very likely both, cannot be trusted \cite{Kurkela:2014vha}.

Previous estimates of the surface tension were either performed in the framework of chiral models that lack the nuclear matter ingredient or employed two different models for nuclear and quark matter, which are
glued together at the phase transition, treating the surface tension as a free parameter \cite{Glendenning:1997wn}. We desire a self-consistent framework where both phases are available at once, ({\emph{e.g.}}, \cite{Jokela:2020piw}) and so can determine the surface tension by following standard computations \cite{Coleman:1977py,Callan:1977pt,Linde:1981zj} extended to the context of deconfinement phase transitions \cite{Palhares:2010be,Fraga:2018cvr}. One of the major goal of our program is to show how the quantum effective action is obtained using gauge/gravity duality and then predicting the surface tension and all other quasi-stationary parameters \cite{shortcompanion} at the deconfinement phase transition. This goal is achieved by extending our work to non-zero chemical potential.

\begin{acknowledgments}
We thank Matti J\"arvinen, Jani Kastikainen, Elias Kiritsis, and Francesco Nitti for useful discussions. 
F.~R.~A. has been supported through the STFC/UKRI grant no.~2131876. 
O.~H. has been supported by the
Academy of Finland grant no.~1330346, the Ruth and Nils-Erik Stenb\"ack foundation, and the Waldemar von Frenckell foundation. 
M.~H. (ORCID ID 0000-0002-9307-437X) acknowledges support from  the Academy of Finland grant no.~333609. 
C.~H. has been partially supported by the Spanish Ministerio de Ciencia, Innovaci\'on y Universidades through the grant PGC2018-096894-B-100. N.~J. has been supported by the Academy of Finland grant no.~1322307.
\end{acknowledgments}


\appendix

\section{Boundary analysis and holographic renormalisation}\label{app:holoRenorm}

It is convenient to define the function
\begin{equation}
 k(r) \equiv r^2 e^{-\chi(r)}\left[rh'(r)-2h(r) - 2rh(r)\chi'(r)\right] \ ,
\end{equation}
since it can be shown using the equations of motion that it is constant, $k'(r)=0$. Evaluating this function on the horizon and on the AdS boundary gives the equality
\begin{equation}
 r_H^3 e^{-\chi(r_H)}h'(r_H) = -4h_2 + \frac{128}{27}\phi_-\phi_+ \ ,
\end{equation}
where we have used the asymptotic solution (\ref{eq:falloff}). This can be rewritten in terms of the temperature and entropy density (\ref{eq:thermo}) as
\begin{equation}
 h_2 = -\frac{\kappa^2_5 T s}{2} + \frac{32}{27}\phi_-\phi_+ \ .
\end{equation}

Turning now to the gravity action (\ref{eq:gravityAction}), we can use the equations of motion to show that on-shell it can be written as a total derivative in the radial coordinate:
\begin{equation}
 S_{bulk}\overset{on-shell}{\longrightarrow}\frac{1}{\kappa^2_5}\int d^{4}x\,dr\left[-\partial_{r}\left(\sqrt{g}\frac{{h(r)}}{r}\right)\right]=-\frac{1}{\kappa^2_5}\int d^{4}x\left[\sqrt{g}\frac{{h(r)}}{r}\right]_{r_{H}}^{r_{\infty}} \ .
\end{equation}
Here we have integrated from the horizon $r_{H}$ to some cutoff surface at a radius $r_{\infty}$. As usual the on-shell action diverges as the cutoff is taken to infinity, requiring renormalisation through the addition of counter-terms defined on the cutoff surface. In the present case, these are
\begin{equation}
S_{CT}=\frac{1}{\kappa^2_5}\int d^{4}x\sqrt{\gamma}\left[c_{0}+c_{1}\phi(r)n^{\mu}\partial_{\mu}\phi(r)+c_{2}\phi(r)^{2}+c_{3}\phi(r)^{3}\right]\ ,
\end{equation}
where $\gamma$ is the determinant of the induced metric $\gamma_{ij}$ on the cutoff surface, and the $c_i$'s are constants to be fixed shortly. Note that the cubic term only gives a finite contribution as the cutoff is taken to infinity. Terms of even higher order vanish at the boundary and are therefore not necessary to include. We also include a Gibbons-Hawking term
\begin{equation}
S_{GH}=\frac{1}{\kappa^2_5}\int d^{4}x\sqrt{\gamma}K \ ,
\end{equation}
where $K$ is the trace of the extrinsic curvature $K_{ij}$.

The complete gravity action including boundary terms is then $S_{C}=S_{bulk}+S_{GH}+S_{CT}$. Requiring that $S_{C}$ is finite as $r_{\infty}\rightarrow\infty$ imposes constraints on the counter-terms:
\begin{equation}
c_{0}=-3\qquad\text{{and}\ensuremath{\qquad}}c_{2}=\frac{{2}}{3}(2c_{1}-1) \ .
\end{equation}
The finite result for the complete action on-shell can then be written in terms of the constants of the near-boundary expansion as
\begin{equation}
S_{C}\overset{on-shell}{\longrightarrow} \frac{\beta V_3}{\kappa^2_5} \left( -\frac{{h_{2}}}{2}+\left(\frac{{28}}{27}-\frac{4}{3}c_{1}\right)\phi_-\phi_++c_{3}\phi_-^{3} \right) \ .
\end{equation}
Here we have also carried out the integration in the Euclidean time direction, giving a factor of $\beta=1/T$, and the three spatial directions, giving a formally infinite volume factor which we denote by $V_3$. If we instead vary our action --- including the counter-terms --- with respect to the scalar field $\phi$, we find the following:
\begin{equation}
\delta_{\phi}S_{C}=\frac{1}{\kappa^2_5}\int d^4x \left\{ -\frac{{4}}{3}c_{1}\phi_-\,\delta\phi_++\left(\frac{{4}}{3}(1-c_{1})\phi_++3c_{3}\phi_-^{2}\right)\delta\phi_- \right\} \ .
\end{equation}
This variation should vanish on solutions, but there are several possible ways to make that happen.

\subsection{Standard quantisation}

In the standard case, the leading falloff $\phi_-$ is fixed. With this choice, the dual CFT has a dimension 8/3 (single trace) operator $\Psi$. A geometry with boundary condition $\phi_-=0$ is dual to an undeformed state of this CFT, and a geometry obeying $\phi_-=\Lambda$ is dual to a state of the deformed CFT $S_{CFT}\rightarrow S_{CFT}+\Lambda \Psi$.

Fixing $\phi_-$ means $\delta\phi_-=0$. To make the variation of the action above vanish on solutions, we are forced to set $c_{1}=0$. The expectation value of the dual operator is then given by
\begin{equation}
\kappa^2_5 \frac{{\delta_{\phi}S_{C}}}{\delta\phi_-}=\frac{{4}}{3}\phi_++3c_{3}\phi_-^{2}\ .
\end{equation}
Note that the constant $c_{3}$ of the finite counter-term is still unfixed, and that the expectation value depends on it. 
The on-shell action becomes
\begin{equation}
S_{C}\overset{on-shell}{\longrightarrow} \frac{\beta V_3}{\kappa^2_5} \left( -\frac{{h_{2}}}{2}+\frac{{28}}{27}\phi_-\phi_++c_{3}\phi_-^{3} \right) \ .
\end{equation}

\subsection{Alternate quantisation}

In the mass range (\ref{eq:MTrange}), one can instead choose to fix the sub-leading falloff $\phi_+$ --- this is the choice we are mainly interested in, since it also allows for multitrace deformations. With this choice, the dual CFT has a dimension 4/3 single trace operator $\Psi$. A geometry with boundary condition $\phi_+=0$ is dual to an undeformed state of this CFT, and a geometry obeying $\phi_+=\Lambda$ is dual to a state of the deformed CFT $S_{CFT}\rightarrow S_{CFT}+\Lambda \Psi$.

Fixing $\phi_+$ means $\delta\phi_+=0$. To make the variation of the action vanish on solutions, we are then forced to set $c_{1}=1$ and $c_{3}=0$. Then the expectation value of the dual operator is given by
\begin{equation}\label{eq:EValtQuant}
\psi \equiv \kappa^2_5\frac{{\delta_{\phi}S_{C}}}{\delta\phi_+}=-\frac{{4}}{3}\phi_-\ .
\end{equation}
The on-shell action becomes
\begin{equation}
S_{C}\overset{on-shell}{\longrightarrow} \frac{\beta V_3}{\kappa^2_5} \left( -\frac{{h_{2}}}{2}-\frac{8}{27}\phi_-\phi_+ \right) \ .
\end{equation}

\subsection{Double and triple trace deformation}

We now consider deforming the alternate quantisation CFT by a double-trace and a triple-trace deformation. This requires the addition of a new, non-covariant counter-term to the action. We thus define the full action to be $S_{C}=S_{bulk}+S_{GH}+S_{CT}+S_{W}$, with
\begin{equation}
S_{W}=\frac{1}{\kappa^2_5}\int d^{4}x\sqrt{g}\left[\psi\, W'(\psi)-W(\psi)\right] \ ,
\end{equation}
and $\psi$ given by (\ref{eq:EValtQuant}). As in the previous subsection, we set $c_{1}=1$. The on-shell action becomes
\begin{equation}
S_{C}\overset{on-shell}{\longrightarrow} \frac{\beta V_3}{\kappa^2_5} \left( -\frac{h_{2}}{2}-\frac{8}{27}\phi_-\phi_++c_{3}\phi_-^{3}+\psi\, W'(\psi)-W(\psi) \right) \ .
\end{equation}
and its variation is
\begin{equation}
\begin{split}
\delta_{\phi}S_{C}&= \frac{1}{\kappa^2_5}\int d^4x \left\{ -\frac{4}{3}\phi_-\,\delta\phi_++3c_{3}\phi_-^{2}\delta\phi_- + \frac{16}{9}\phi_-\, W''(\psi)\delta\phi_- \right\} \\
&= \frac{1}{\kappa^2_5}\int d^4x \left\{ -\frac{4}{3}\phi_-\delta\left(\phi_+-\frac{9}{8}c_{3}\phi_-^{2} + W'(\psi)\right) \right\} \ .
\end{split}
\end{equation}
We can see that in this setup, the constant $c_3$ simply shifts the cubic term in $W$, so we will set $c_{3}=0$ and instead let
\begin{equation}
W(\psi)=\frac{f}{2}\psi^{2}+\frac{g}{3}\psi^{3} 
\end{equation}
giving
\begin{equation}\label{eq:OSactionGeneral}
S_{C}\overset{on-shell}{\longrightarrow} \frac{\beta V_3}{\kappa^2_5} \left( -\frac{h_{2}}{2}-\frac{8}{27}\phi_-\phi_+ + \frac{8}{9}f\phi_-^{2} - \frac{128}{81}g\phi_-^{3}\right) 
\end{equation}
and
\begin{equation}
\delta_{\phi}S_{C}=\frac{1}{\kappa^2_5}\int d^4x \left\{ -\frac{4}{3}\phi_-\delta\left(\phi_+-\frac{4}{3}f\phi_-+\frac{16}{9}g\phi_-^{2}\right)\right\} \ .
\end{equation}
The variation now vanishes on solutions with the boundary condition
\begin{equation}\label{eq:BCgeneral}
\phi_+ - \frac{4}{3}f\phi_- + \frac{16}{9}g\phi_-^{2}=J 
\end{equation}
with $J$ a constant. This very general boundary condition allows for a single-, double-, and triple-trace deformation of the original CFT.

The field theory generating functional ${\cal W}[J]$ is equal to the gravitational on-shell action. For uniform fields and sources, we define $w(J)\equiv \kappa^2_5 {\cal W}[J]/(\beta V_3)$; then we can write
\begin{equation}\label{eq:generatingFunc}
 w(J) = -\frac{h_{2}(J)}{2}-\frac{8}{27}\phi_-(J)\, J + \frac{40}{81}f\phi_-^{2}(J) - \frac{256}{243}g\phi_-^{3}(J) \ .
\end{equation}
Here we have used (\ref{eq:BCgeneral}), and we emphasize that in this expression, $h_2$ and $\phi_-$ should be regarded as functions of $J$, which we need to solve the full gravitational equations of motion to determine.

To get an expression for the effective potential, we substitute (\ref{eq:EValtQuant}), (\ref{eq:BCgeneral}), and (\ref{eq:generatingFunc}) into (\ref{eq:effPotUniform}), giving
\begin{equation}
 V(\psi) = -w(J)+\psi J = \frac{h_{2}(\psi)}{2} + \frac{7}{9}\phi_+(\psi)\, \psi + \frac{f}{2}\psi^2 + \frac{g}{3}\psi^3  \ .
\end{equation}
Note that all the non-trivial information contained within the coefficients $h_2$ and $\phi_+$, which as we indicate should now be regarded as functions of $\psi=-\frac{4}{3}\phi_-$; these must be extracted from numerical solutions. Meanwhile, the multi-trace deformations give simple polynomial contributions. We can furthermore include also the possibility of a single-trace deformation $\Lambda\Psi$ by shifting $J\to J-\Lambda$ in (\ref{eq:BCgeneral}) before substituting it into (\ref{eq:effPotUniform}), giving
\begin{equation}\label{eq:effPotCompleteA}
 V(\psi) = \frac{h_{2}(\psi)}{2} + \frac{7}{9}\phi_+(\psi)\, \psi + \Lambda \psi + \frac{f}{2}\psi^2 + \frac{g}{3}\psi^3  \ .
\end{equation}

\section{Exact result for the effective potential at large temperatures}\label{app:effPotV2}

As explained in Sec.~\ref{sec:effActionHolography}, the second derivative of the effective action gives the inverse of the two-point function. In the gravitational bulk, two-point functions can be computed by a fluctuation analysis. For our particular holographic model, we derived (\ref{eq:Gamma2}), which we reproduce here:
\begin{equation}\label{eq:Gamma2A}
 \Gamma_2 = -\left( \frac{3}{4}\frac{Z_\phi^{+(0)}}{Z_\phi^{-(0)}} + W''(v) \right) + \frac{3}{4}\frac{Z_\phi^{-(2)}Z_\phi^{+(0)}-Z_\phi^{-(0)}Z_\phi^{+(2)}}{\left(Z_\phi^{-(0)}\right)^2}k^2 + \ldots \ .
\end{equation}
Here, the first term of order $k^0$ should equal the second derivative of the effective potential.

In general we can only determine these terms numerically. However, in the limit of small field values --- or equivalently, large temperatures --- the background approaches pure AdS-Schwarzschild, where it is possible to find an analytic solution. In this background, the two gauge invariant modes in (\ref{eq:gaugeInvModes}) decouple, and the equation for $Z_\phi(r)$ with $k=0$ takes the form
\begin{equation}
 Z_\phi^{(0)\prime\prime}(z)-\frac{3+z^4}{z-z^5}Z_\phi^{(0)\prime}(z) + \frac{32}{9}\frac{Z_\phi^{(0)}(z)}{z^2-z^6}=0 \ ,
\end{equation}
where we have switched to the radial coordinate $z=r_H/r$. This can be solved in terms of hypergeometric functions as
\begin{equation}
 Z_\phi^{(0)}(z) = c_1 z^{4/3}\, \prescript{}{2}{F}_1\left[ 1/3 , 1/3 , 2/3 , z^4 \right] + c_2 z^{8/3}\, \prescript{}{2}{F}_1\left[ 2/3 , 2/3 , 4/3 , z^4 \right] \ .
\end{equation}
Regularity on the horizon $z=1$ imposes 
\begin{equation}
 \frac{c_2}{c_1} = -\frac{\Gamma(2/3)^3}{\Gamma(1/3)^2\, \Gamma(4/3)} \ .
\end{equation}
Expanding the result near the AdS boundary at $z=0$, we then find
\begin{equation}
 \frac{Z_\phi^{+(0)}}{Z_\phi^{-(0)}} = -\frac{\pi^{3/2}r_H^{4/3}}{18\Gamma(7/6)^3} \ .
\end{equation}
Plugging into (\ref{eq:Gamma2A}) and expressing the result in units of temperature, where for AdS-Schwarzschild $T=r_H/\pi$, this provides the coefficient $V_2$ in the small-field (or high-T) expansion of the effective potential (\ref{eq:smallPsiExpansion}). The result is
\begin{equation}
V_2 = \frac{9\pi^{17/6}}{\Gamma(1/6)^3} \ ,
\end{equation}
which is what we quote in (\ref{eq:V2}). This also agrees with the analysis in \cite{Faulkner:2010gj}.

\section{Gauge invariant fluctuation equations}\label{app:fluctuationEquations}

The gauge invariant modes introduced in (\ref{eq:gaugeInvModes}) satisfy the two coupled linear second order differential equations,
\begin{align}
 Z_\phi''(r) + A Z_\phi'(r) + B Z_H'(r) + C Z_\phi(r) + D Z_H(r) &=0 \\
 Z_H''(r) + E Z_H'(r) + F Z_\phi'(r) + G Z_H(r) + H Z_\phi(r) &=0 \ ,
\end{align}
with
\begin{align}
 A =& \frac{1}{r}-\frac{r\mathcal{P}(\phi)}{3h(r)} \\
 B =& \frac{e^{2\chi(r)} r \left(3\mathcal{P}'(\phi)+2r\mathcal{P}(\phi)\phi'(r)\right)}{h(r)\theta(r)} \\
 C =& -\frac{4r^2 \phi'(r)^2 \mathcal{P}(\phi)}{\theta(r)} - \frac{6 r \phi'(r) \mathcal{P}'(\phi)}{\theta(r)} - \frac{k^2}{r^2 h(r)} - \frac{r \phi '(r) \mathcal{P}'(\phi)+3 r^2 \mathcal{P}''(\phi)}{3h(r)} \\
 D =& \frac{r^2 e^{2 \chi (r)} \left(3 \mathcal{P}'(\phi)+2 r
   \mathcal{P}(\phi) \phi '(r)\right) \left(\mathcal{P}(\phi
   )-h(r) \phi '(r)^2\right)}{6 h(r)^2\, \theta(r)} \\
 E =& \frac{r \mathcal{P}(\phi)}{3} \left(\frac{36}{\theta(r)}-\frac{1}{h(r)}\right)-\frac{2}{3}r \phi '(r)^2 +\frac{5}{r} \\
 F =& \ 0 \\
 G =& -\frac{9 k^2+r^4 \mathcal{P}(\phi) \phi
   '(r)^2}{9r^2 h(r)}+\frac{12 \mathcal{P}(\phi)}{\theta(r)}+\frac{4}{r^2}+\frac{r^2 \phi '(r)^4}{9}-\frac{4}{3} \phi
   '(r)^2 \\
 H =& \frac{e^{-2 \chi (r)}}{9 h(r) \theta(r)} \left( \theta(r) - 18h(r)\right) \left( 12 r h(r) \mathcal{P}(\phi) \phi '(r) + \theta(r) \mathcal{P}'(\phi)\right) \ ,
\end{align}
and where we defined
\begin{equation}
 \theta(r)\equiv h(r) \left(r^2 \phi'(r)^2+6\right)-r^2 \mathcal{P}(\phi) \ .
\end{equation}

\bibliographystyle{apsrev4-2}

\bibliography{biblio}

\begin{thebibliography}{50}%
\makeatletter
\providecommand \@ifxundefined [1]{%
 \@ifx{#1\undefined}
}%
\providecommand \@ifnum [1]{%
 \ifnum #1\expandafter \@firstoftwo
 \else \expandafter \@secondoftwo
 \fi
}%
\providecommand \@ifx [1]{%
 \ifx #1\expandafter \@firstoftwo
 \else \expandafter \@secondoftwo
 \fi
}%
\providecommand \natexlab [1]{#1}%
\providecommand \enquote  [1]{``#1''}%
\providecommand \bibnamefont  [1]{#1}%
\providecommand \bibfnamefont [1]{#1}%
\providecommand \citenamefont [1]{#1}%
\providecommand \href@noop [0]{\@secondoftwo}%
\providecommand \href [0]{\begingroup \@sanitize@url \@href}%
\providecommand \@href[1]{\@@startlink{#1}\@@href}%
\providecommand \@@href[1]{\endgroup#1\@@endlink}%
\providecommand \@sanitize@url [0]{\catcode `\\12\catcode `\$12\catcode
  `\&12\catcode `\#12\catcode `\^12\catcode `\_12\catcode `\%12\relax}%
\providecommand \@@startlink[1]{}%
\providecommand \@@endlink[0]{}%
\providecommand \url  [0]{\begingroup\@sanitize@url \@url }%
\providecommand \@url [1]{\endgroup\@href {#1}{\urlprefix }}%
\providecommand \urlprefix  [0]{URL }%
\providecommand \Eprint [0]{\href }%
\providecommand \doibase [0]{http://dx.doi.org/}%
\providecommand \selectlanguage [0]{\@gobble}%
\providecommand \bibinfo  [0]{\@secondoftwo}%
\providecommand \bibfield  [0]{\@secondoftwo}%
\providecommand \translation [1]{[#1]}%
\providecommand \BibitemOpen [0]{}%
\providecommand \bibitemStop [0]{}%
\providecommand \bibitemNoStop [0]{.\EOS\space}%
\providecommand \EOS [0]{\spacefactor3000\relax}%
\providecommand \BibitemShut  [1]{\csname bibitem#1\endcsname}%
\let\auto@bib@innerbib\@empty
\bibitem [{\citenamefont {Amaro-Seoane}\ \emph {et~al.}(2017)\citenamefont
  {Amaro-Seoane} \emph {et~al.}}]{LISA:2017pwj}%
  \BibitemOpen
  \bibfield  {author} {\bibinfo {author} {\bibfnamefont {P.}~\bibnamefont
  {Amaro-Seoane}} \emph {et~al.} (\bibinfo {collaboration} {LISA}),\
  }\href@noop {} {\  (\bibinfo {year} {2017})},\ \Eprint
  {http://arxiv.org/abs/1702.00786} {arXiv:1702.00786 [astro-ph.IM]}
  \BibitemShut {NoStop}%
\bibitem [{\citenamefont {Caprini}\ \emph {et~al.}(2020)\citenamefont {Caprini}
  \emph {et~al.}}]{Caprini:2019egz}%
  \BibitemOpen
  \bibfield  {author} {\bibinfo {author} {\bibfnamefont {C.}~\bibnamefont
  {Caprini}} \emph {et~al.},\ }\href {\doibase 10.1088/1475-7516/2020/03/024}
  {\bibfield  {journal} {\bibinfo  {journal} {JCAP}\ }\textbf {\bibinfo
  {volume} {03}},\ \bibinfo {pages} {024} (\bibinfo {year} {2020})},\ \Eprint
  {http://arxiv.org/abs/1910.13125} {arXiv:1910.13125 [astro-ph.CO]}
  \BibitemShut {NoStop}%
\bibitem [{\citenamefont {Mazumdar}\ and\ \citenamefont
  {White}(2019)}]{Mazumdar:2018dfl}%
  \BibitemOpen
  \bibfield  {author} {\bibinfo {author} {\bibfnamefont {A.}~\bibnamefont
  {Mazumdar}}\ and\ \bibinfo {author} {\bibfnamefont {G.}~\bibnamefont
  {White}},\ }\href {\doibase 10.1088/1361-6633/ab1f55} {\bibfield  {journal}
  {\bibinfo  {journal} {Rept. Prog. Phys.}\ }\textbf {\bibinfo {volume} {82}},\
  \bibinfo {pages} {076901} (\bibinfo {year} {2019})},\ \Eprint
  {http://arxiv.org/abs/1811.01948} {arXiv:1811.01948 [hep-ph]} \BibitemShut
  {NoStop}%
\bibitem [{\citenamefont {Hindmarsh}\ \emph {et~al.}(2020)\citenamefont
  {Hindmarsh}, \citenamefont {L{\"u}ben}, \citenamefont {Lumma},\ and\
  \citenamefont {Pauly}}]{Hindmarsh:2020hop}%
  \BibitemOpen
  \bibfield  {author} {\bibinfo {author} {\bibfnamefont {M.~B.}\ \bibnamefont
  {Hindmarsh}}, \bibinfo {author} {\bibfnamefont {M.}~\bibnamefont
  {L{\"u}ben}}, \bibinfo {author} {\bibfnamefont {J.}~\bibnamefont {Lumma}}, \
  and\ \bibinfo {author} {\bibfnamefont {M.}~\bibnamefont {Pauly}},\
  }\href@noop {} {\  (\bibinfo {year} {2020})},\ \Eprint
  {http://arxiv.org/abs/2008.09136} {arXiv:2008.09136 [astro-ph.CO]}
  \BibitemShut {NoStop}%
\bibitem [{\citenamefont {Halverson}\ \emph {et~al.}(2021)\citenamefont
  {Halverson}, \citenamefont {Long}, \citenamefont {Maiti}, \citenamefont
  {Nelson},\ and\ \citenamefont {Salinas}}]{Halverson:2020xpg}%
  \BibitemOpen
  \bibfield  {author} {\bibinfo {author} {\bibfnamefont {J.}~\bibnamefont
  {Halverson}}, \bibinfo {author} {\bibfnamefont {C.}~\bibnamefont {Long}},
  \bibinfo {author} {\bibfnamefont {A.}~\bibnamefont {Maiti}}, \bibinfo
  {author} {\bibfnamefont {B.}~\bibnamefont {Nelson}}, \ and\ \bibinfo {author}
  {\bibfnamefont {G.}~\bibnamefont {Salinas}},\ }\href {\doibase
  10.1007/JHEP05(2021)154} {\bibfield  {journal} {\bibinfo  {journal} {JHEP}\
  }\textbf {\bibinfo {volume} {05}},\ \bibinfo {pages} {154} (\bibinfo {year}
  {2021})},\ \Eprint {http://arxiv.org/abs/2012.04071} {arXiv:2012.04071
  [hep-ph]} \BibitemShut {NoStop}%
\bibitem [{\citenamefont {Huang}\ \emph {et~al.}(2021)\citenamefont {Huang},
  \citenamefont {Reichert}, \citenamefont {Sannino},\ and\ \citenamefont
  {Wang}}]{Huang:2020crf}%
  \BibitemOpen
  \bibfield  {author} {\bibinfo {author} {\bibfnamefont {W.-C.}\ \bibnamefont
  {Huang}}, \bibinfo {author} {\bibfnamefont {M.}~\bibnamefont {Reichert}},
  \bibinfo {author} {\bibfnamefont {F.}~\bibnamefont {Sannino}}, \ and\
  \bibinfo {author} {\bibfnamefont {Z.-W.}\ \bibnamefont {Wang}},\ }\href
  {\doibase 10.1103/PhysRevD.104.035005} {\bibfield  {journal} {\bibinfo
  {journal} {Phys. Rev. D}\ }\textbf {\bibinfo {volume} {104}},\ \bibinfo
  {pages} {035005} (\bibinfo {year} {2021})},\ \Eprint
  {http://arxiv.org/abs/2012.11614} {arXiv:2012.11614 [hep-ph]} \BibitemShut
  {NoStop}%
\bibitem [{\citenamefont {Reichert}\ \emph {et~al.}(2021)\citenamefont
  {Reichert}, \citenamefont {Sannino}, \citenamefont {Wang},\ and\
  \citenamefont {Zhang}}]{Reichert:2021cvs}%
  \BibitemOpen
  \bibfield  {author} {\bibinfo {author} {\bibfnamefont {M.}~\bibnamefont
  {Reichert}}, \bibinfo {author} {\bibfnamefont {F.}~\bibnamefont {Sannino}},
  \bibinfo {author} {\bibfnamefont {Z.-W.}\ \bibnamefont {Wang}}, \ and\
  \bibinfo {author} {\bibfnamefont {C.}~\bibnamefont {Zhang}},\ }\href@noop {}
  {\  (\bibinfo {year} {2021})},\ \Eprint {http://arxiv.org/abs/2109.11552}
  {arXiv:2109.11552 [hep-ph]} \BibitemShut {NoStop}%
\bibitem [{\citenamefont {Bigazzi}\ \emph
  {et~al.}(2021{\natexlab{a}})\citenamefont {Bigazzi}, \citenamefont {Caddeo},
  \citenamefont {Cotrone},\ and\ \citenamefont {Paredes}}]{Bigazzi:2020avc}%
  \BibitemOpen
  \bibfield  {author} {\bibinfo {author} {\bibfnamefont {F.}~\bibnamefont
  {Bigazzi}}, \bibinfo {author} {\bibfnamefont {A.}~\bibnamefont {Caddeo}},
  \bibinfo {author} {\bibfnamefont {A.~L.}\ \bibnamefont {Cotrone}}, \ and\
  \bibinfo {author} {\bibfnamefont {A.}~\bibnamefont {Paredes}},\ }\href
  {\doibase 10.1007/JHEP04(2021)094} {\bibfield  {journal} {\bibinfo  {journal}
  {JHEP}\ }\textbf {\bibinfo {volume} {04}},\ \bibinfo {pages} {094} (\bibinfo
  {year} {2021}{\natexlab{a}})},\ \Eprint {http://arxiv.org/abs/2011.08757}
  {arXiv:2011.08757 [hep-ph]} \BibitemShut {NoStop}%
\bibitem [{\citenamefont {Ares}\ \emph {et~al.}(2020)\citenamefont {Ares},
  \citenamefont {Hindmarsh}, \citenamefont {Hoyos},\ and\ \citenamefont
  {Jokela}}]{Ares:2020lbt}%
  \BibitemOpen
  \bibfield  {author} {\bibinfo {author} {\bibfnamefont {F.~R.}\ \bibnamefont
  {Ares}}, \bibinfo {author} {\bibfnamefont {M.}~\bibnamefont {Hindmarsh}},
  \bibinfo {author} {\bibfnamefont {C.}~\bibnamefont {Hoyos}}, \ and\ \bibinfo
  {author} {\bibfnamefont {N.}~\bibnamefont {Jokela}},\ }\href {\doibase
  10.1007/JHEP04(2021)100} {\bibfield  {journal} {\bibinfo  {journal} {JHEP}\
  }\textbf {\bibinfo {volume} {21}},\ \bibinfo {pages} {100} (\bibinfo {year}
  {2020})},\ \Eprint {http://arxiv.org/abs/2011.12878} {arXiv:2011.12878
  [hep-th]} \BibitemShut {NoStop}%
\bibitem [{\citenamefont {Zhu}\ \emph {et~al.}(2021)\citenamefont {Zhu},
  \citenamefont {Chen},\ and\ \citenamefont {Hou}}]{Zhu:2021vkj}%
  \BibitemOpen
  \bibfield  {author} {\bibinfo {author} {\bibfnamefont {Z.-R.}\ \bibnamefont
  {Zhu}}, \bibinfo {author} {\bibfnamefont {J.}~\bibnamefont {Chen}}, \ and\
  \bibinfo {author} {\bibfnamefont {D.}~\bibnamefont {Hou}},\ }\href@noop {} {\
   (\bibinfo {year} {2021})},\ \Eprint {http://arxiv.org/abs/2109.09933}
  {arXiv:2109.09933 [hep-ph]} \BibitemShut {NoStop}%
\bibitem [{\citenamefont {Bigazzi}\ \emph {et~al.}(2020)\citenamefont
  {Bigazzi}, \citenamefont {Caddeo}, \citenamefont {Cotrone},\ and\
  \citenamefont {Paredes}}]{Bigazzi:2020phm}%
  \BibitemOpen
  \bibfield  {author} {\bibinfo {author} {\bibfnamefont {F.}~\bibnamefont
  {Bigazzi}}, \bibinfo {author} {\bibfnamefont {A.}~\bibnamefont {Caddeo}},
  \bibinfo {author} {\bibfnamefont {A.~L.}\ \bibnamefont {Cotrone}}, \ and\
  \bibinfo {author} {\bibfnamefont {A.}~\bibnamefont {Paredes}},\ }\href
  {\doibase 10.1007/JHEP12(2020)200} {\bibfield  {journal} {\bibinfo  {journal}
  {JHEP}\ }\textbf {\bibinfo {volume} {12}},\ \bibinfo {pages} {200} (\bibinfo
  {year} {2020})},\ \Eprint {http://arxiv.org/abs/2008.02579} {arXiv:2008.02579
  [hep-th]} \BibitemShut {NoStop}%
\bibitem [{\citenamefont {Bea}\ \emph {et~al.}(2021)\citenamefont {Bea},
  \citenamefont {Casalderrey-Solana}, \citenamefont {Giannakopoulos},
  \citenamefont {Mateos}, \citenamefont {Sanchez-Garitaonandia},\ and\
  \citenamefont {Zilh{\~a}o}}]{Bea:2021zsu}%
  \BibitemOpen
  \bibfield  {author} {\bibinfo {author} {\bibfnamefont {Y.}~\bibnamefont
  {Bea}}, \bibinfo {author} {\bibfnamefont {J.}~\bibnamefont
  {Casalderrey-Solana}}, \bibinfo {author} {\bibfnamefont {T.}~\bibnamefont
  {Giannakopoulos}}, \bibinfo {author} {\bibfnamefont {D.}~\bibnamefont
  {Mateos}}, \bibinfo {author} {\bibfnamefont {M.}~\bibnamefont
  {Sanchez-Garitaonandia}}, \ and\ \bibinfo {author} {\bibfnamefont
  {M.}~\bibnamefont {Zilh{\~a}o}},\ }\href@noop {} {\  (\bibinfo {year}
  {2021})},\ \Eprint {http://arxiv.org/abs/2104.05708} {arXiv:2104.05708
  [hep-th]} \BibitemShut {NoStop}%
\bibitem [{\citenamefont {Bigazzi}\ \emph
  {et~al.}(2021{\natexlab{b}})\citenamefont {Bigazzi}, \citenamefont {Caddeo},
  \citenamefont {Canneti},\ and\ \citenamefont {Cotrone}}]{Bigazzi:2021fmq}%
  \BibitemOpen
  \bibfield  {author} {\bibinfo {author} {\bibfnamefont {F.}~\bibnamefont
  {Bigazzi}}, \bibinfo {author} {\bibfnamefont {A.}~\bibnamefont {Caddeo}},
  \bibinfo {author} {\bibfnamefont {T.}~\bibnamefont {Canneti}}, \ and\
  \bibinfo {author} {\bibfnamefont {A.~L.}\ \bibnamefont {Cotrone}},\ }\href
  {\doibase 10.1007/JHEP08(2021)090} {\  (\bibinfo {year}
  {2021}{\natexlab{b}}),\ 10.1007/JHEP08(2021)090},\ \Eprint
  {http://arxiv.org/abs/2104.12817} {arXiv:2104.12817 [hep-ph]} \BibitemShut
  {NoStop}%
\bibitem [{\citenamefont {Henriksson}(2021)}]{Henriksson:2021zei}%
  \BibitemOpen
  \bibfield  {author} {\bibinfo {author} {\bibfnamefont {O.}~\bibnamefont
  {Henriksson}},\ }\href@noop {} {\  (\bibinfo {year} {2021})},\ \Eprint
  {http://arxiv.org/abs/2106.13254} {arXiv:2106.13254 [hep-th]} \BibitemShut
  {NoStop}%
\bibitem [{\citenamefont {Gould}\ and\ \citenamefont
  {Tenkanen}(2021)}]{Gould:2021oba}%
  \BibitemOpen
  \bibfield  {author} {\bibinfo {author} {\bibfnamefont {O.}~\bibnamefont
  {Gould}}\ and\ \bibinfo {author} {\bibfnamefont {T.~V.~I.}\ \bibnamefont
  {Tenkanen}},\ }\href {\doibase 10.1007/JHEP06(2021)069} {\bibfield  {journal}
  {\bibinfo  {journal} {JHEP}\ }\textbf {\bibinfo {volume} {06}},\ \bibinfo
  {pages} {069} (\bibinfo {year} {2021})},\ \Eprint
  {http://arxiv.org/abs/2104.04399} {arXiv:2104.04399 [hep-ph]} \BibitemShut
  {NoStop}%
\bibitem [{\citenamefont {Moore}\ and\ \citenamefont
  {Rummukainen}(2001)}]{Moore:2000jw}%
  \BibitemOpen
  \bibfield  {author} {\bibinfo {author} {\bibfnamefont {G.~D.}\ \bibnamefont
  {Moore}}\ and\ \bibinfo {author} {\bibfnamefont {K.}~\bibnamefont
  {Rummukainen}},\ }\href {\doibase 10.1103/PhysRevD.63.045002} {\bibfield
  {journal} {\bibinfo  {journal} {Phys. Rev. D}\ }\textbf {\bibinfo {volume}
  {63}},\ \bibinfo {pages} {045002} (\bibinfo {year} {2001})},\ \Eprint
  {http://arxiv.org/abs/hep-ph/0009132} {arXiv:hep-ph/0009132} \BibitemShut
  {NoStop}%
\bibitem [{\citenamefont {Ares}\ \emph {et~al.}(2021)\citenamefont {Ares},
  \citenamefont {Henriksson}, \citenamefont {Hindmarsh}, \citenamefont
  {Hoyos},\ and\ \citenamefont {Jokela}}]{shortcompanion}%
  \BibitemOpen
  \bibfield  {author} {\bibinfo {author} {\bibfnamefont {F.~R.}\ \bibnamefont
  {Ares}}, \bibinfo {author} {\bibfnamefont {O.}~\bibnamefont {Henriksson}},
  \bibinfo {author} {\bibfnamefont {M.}~\bibnamefont {Hindmarsh}}, \bibinfo
  {author} {\bibfnamefont {C.}~\bibnamefont {Hoyos}}, \ and\ \bibinfo {author}
  {\bibfnamefont {N.}~\bibnamefont {Jokela}},\ }\href@noop {} {\  (\bibinfo
  {year} {2021})},\ \Eprint {http://arxiv.org/abs/2110.14442} {arXiv:2110.14442
  [hep-th]} \BibitemShut {NoStop}%
\bibitem [{\citenamefont {Janik}\ \emph {et~al.}(2021)\citenamefont {Janik},
  \citenamefont {J{\"a}rvinen},\ and\ \citenamefont
  {Sonnenschein}}]{Janik:2021jbq}%
  \BibitemOpen
  \bibfield  {author} {\bibinfo {author} {\bibfnamefont {R.~A.}\ \bibnamefont
  {Janik}}, \bibinfo {author} {\bibfnamefont {M.}~\bibnamefont {J{\"a}rvinen}},
  \ and\ \bibinfo {author} {\bibfnamefont {J.}~\bibnamefont {Sonnenschein}},\
  }\href@noop {} {\  (\bibinfo {year} {2021})},\ \Eprint
  {http://arxiv.org/abs/2106.02642} {arXiv:2106.02642 [hep-th]} \BibitemShut
  {NoStop}%
\bibitem [{\citenamefont {Hertog}\ and\ \citenamefont
  {Horowitz}(2005{\natexlab{a}})}]{Hertog:2004ns}%
  \BibitemOpen
  \bibfield  {author} {\bibinfo {author} {\bibfnamefont {T.}~\bibnamefont
  {Hertog}}\ and\ \bibinfo {author} {\bibfnamefont {G.~T.}\ \bibnamefont
  {Horowitz}},\ }\href {\doibase 10.1103/PhysRevLett.94.221301} {\bibfield
  {journal} {\bibinfo  {journal} {Phys. Rev. Lett.}\ }\textbf {\bibinfo
  {volume} {94}},\ \bibinfo {pages} {221301} (\bibinfo {year}
  {2005}{\natexlab{a}})},\ \Eprint {http://arxiv.org/abs/hep-th/0412169}
  {arXiv:hep-th/0412169} \BibitemShut {NoStop}%
\bibitem [{\citenamefont {Hertog}\ and\ \citenamefont
  {Horowitz}(2005{\natexlab{b}})}]{Hertog:2005hu}%
  \BibitemOpen
  \bibfield  {author} {\bibinfo {author} {\bibfnamefont {T.}~\bibnamefont
  {Hertog}}\ and\ \bibinfo {author} {\bibfnamefont {G.~T.}\ \bibnamefont
  {Horowitz}},\ }\href {\doibase 10.1088/1126-6708/2005/04/005} {\bibfield
  {journal} {\bibinfo  {journal} {JHEP}\ }\textbf {\bibinfo {volume} {04}},\
  \bibinfo {pages} {005} (\bibinfo {year} {2005}{\natexlab{b}})},\ \Eprint
  {http://arxiv.org/abs/hep-th/0503071} {arXiv:hep-th/0503071} \BibitemShut
  {NoStop}%
\bibitem [{\citenamefont {Papadimitriou}(2007)}]{Papadimitriou:2007sj}%
  \BibitemOpen
  \bibfield  {author} {\bibinfo {author} {\bibfnamefont {I.}~\bibnamefont
  {Papadimitriou}},\ }\href {\doibase 10.1088/1126-6708/2007/05/075} {\bibfield
   {journal} {\bibinfo  {journal} {JHEP}\ }\textbf {\bibinfo {volume} {05}},\
  \bibinfo {pages} {075} (\bibinfo {year} {2007})},\ \Eprint
  {http://arxiv.org/abs/hep-th/0703152} {arXiv:hep-th/0703152} \BibitemShut
  {NoStop}%
\bibitem [{\citenamefont {Kiritsis}\ and\ \citenamefont
  {Niarchos}(2012)}]{Kiritsis:2012ma}%
  \BibitemOpen
  \bibfield  {author} {\bibinfo {author} {\bibfnamefont {E.}~\bibnamefont
  {Kiritsis}}\ and\ \bibinfo {author} {\bibfnamefont {V.}~\bibnamefont
  {Niarchos}},\ }\href {\doibase 10.1007/JHEP08(2012)164} {\bibfield  {journal}
  {\bibinfo  {journal} {JHEP}\ }\textbf {\bibinfo {volume} {08}},\ \bibinfo
  {pages} {164} (\bibinfo {year} {2012})},\ \Eprint
  {http://arxiv.org/abs/1205.6205} {arXiv:1205.6205 [hep-th]} \BibitemShut
  {NoStop}%
\bibitem [{\citenamefont {Kiritsis}\ \emph {et~al.}(2014)\citenamefont
  {Kiritsis}, \citenamefont {Li},\ and\ \citenamefont
  {Nitti}}]{Kiritsis:2014kua}%
  \BibitemOpen
  \bibfield  {author} {\bibinfo {author} {\bibfnamefont {E.}~\bibnamefont
  {Kiritsis}}, \bibinfo {author} {\bibfnamefont {W.}~\bibnamefont {Li}}, \ and\
  \bibinfo {author} {\bibfnamefont {F.}~\bibnamefont {Nitti}},\ }\href
  {\doibase 10.1002/prop.201400007} {\bibfield  {journal} {\bibinfo  {journal}
  {Fortsch. Phys.}\ }\textbf {\bibinfo {volume} {62}},\ \bibinfo {pages} {389}
  (\bibinfo {year} {2014})},\ \Eprint {http://arxiv.org/abs/1401.0888}
  {arXiv:1401.0888 [hep-th]} \BibitemShut {NoStop}%
\bibitem [{\citenamefont {Witten}(2001)}]{Witten:2001ua}%
  \BibitemOpen
  \bibfield  {author} {\bibinfo {author} {\bibfnamefont {E.}~\bibnamefont
  {Witten}},\ }\href@noop {} {\  (\bibinfo {year} {2001})},\ \Eprint
  {http://arxiv.org/abs/hep-th/0112258} {arXiv:hep-th/0112258} \BibitemShut
  {NoStop}%
\bibitem [{Note1()}]{Note1}%
  \BibitemOpen
  \bibinfo {note} {The paper \cite {Papadimitriou:2007sj} discusses a class of
  potentials, dubbed the '2/3' potential, which has the same mass plus higher
  order terms; with four instead of five bulk dimensions, this potential can be
  embedded in $\protect \mathcal {N}=8$ gauged supergravity.}\BibitemShut
  {Stop}%
\bibitem [{\citenamefont {Faulkner}\ \emph {et~al.}(2011)\citenamefont
  {Faulkner}, \citenamefont {Horowitz},\ and\ \citenamefont
  {Roberts}}]{Faulkner:2010gj}%
  \BibitemOpen
  \bibfield  {author} {\bibinfo {author} {\bibfnamefont {T.}~\bibnamefont
  {Faulkner}}, \bibinfo {author} {\bibfnamefont {G.~T.}\ \bibnamefont
  {Horowitz}}, \ and\ \bibinfo {author} {\bibfnamefont {M.~M.}\ \bibnamefont
  {Roberts}},\ }\href {\doibase 10.1007/JHEP04(2011)051} {\bibfield  {journal}
  {\bibinfo  {journal} {JHEP}\ }\textbf {\bibinfo {volume} {04}},\ \bibinfo
  {pages} {051} (\bibinfo {year} {2011})},\ \Eprint
  {http://arxiv.org/abs/1008.1581} {arXiv:1008.1581 [hep-th]} \BibitemShut
  {NoStop}%
\bibitem [{\citenamefont {Coleman}(1977)}]{Coleman:1977py}%
  \BibitemOpen
  \bibfield  {author} {\bibinfo {author} {\bibfnamefont {S.~R.}\ \bibnamefont
  {Coleman}},\ }\href {\doibase 10.1103/PhysRevD.16.1248} {\bibfield  {journal}
  {\bibinfo  {journal} {Phys. Rev. D}\ }\textbf {\bibinfo {volume} {15}},\
  \bibinfo {pages} {2929} (\bibinfo {year} {1977})},\ \bibinfo {note}
  {[Erratum: Phys.Rev.D 16, 1248 (1977)]}\BibitemShut {NoStop}%
\bibitem [{\citenamefont {Callan}\ and\ \citenamefont
  {Coleman}(1977)}]{Callan:1977pt}%
  \BibitemOpen
  \bibfield  {author} {\bibinfo {author} {\bibfnamefont {C.~G.}\ \bibnamefont
  {Callan}, \bibfnamefont {Jr.}}\ and\ \bibinfo {author} {\bibfnamefont
  {S.~R.}\ \bibnamefont {Coleman}},\ }\href {\doibase 10.1103/PhysRevD.16.1762}
  {\bibfield  {journal} {\bibinfo  {journal} {Phys. Rev. D}\ }\textbf {\bibinfo
  {volume} {16}},\ \bibinfo {pages} {1762} (\bibinfo {year}
  {1977})}\BibitemShut {NoStop}%
\bibitem [{\citenamefont {Linde}(1983)}]{Linde:1981zj}%
  \BibitemOpen
  \bibfield  {author} {\bibinfo {author} {\bibfnamefont {A.~D.}\ \bibnamefont
  {Linde}},\ }\href {\doibase 10.1016/0550-3213(83)90072-X} {\bibfield
  {journal} {\bibinfo  {journal} {Nucl. Phys. B}\ }\textbf {\bibinfo {volume}
  {216}},\ \bibinfo {pages} {421} (\bibinfo {year} {1983})},\ \bibinfo {note}
  {[Erratum: Nucl.Phys.B 223, 544 (1983)]}\BibitemShut {NoStop}%
\bibitem [{\citenamefont {Lucini}\ \emph {et~al.}(2004)\citenamefont {Lucini},
  \citenamefont {Teper},\ and\ \citenamefont {Wenger}}]{Lucini:2003zr}%
  \BibitemOpen
  \bibfield  {author} {\bibinfo {author} {\bibfnamefont {B.}~\bibnamefont
  {Lucini}}, \bibinfo {author} {\bibfnamefont {M.}~\bibnamefont {Teper}}, \
  and\ \bibinfo {author} {\bibfnamefont {U.}~\bibnamefont {Wenger}},\ }\href
  {\doibase 10.1088/1126-6708/2004/01/061} {\bibfield  {journal} {\bibinfo
  {journal} {JHEP}\ }\textbf {\bibinfo {volume} {01}},\ \bibinfo {pages} {061}
  (\bibinfo {year} {2004})},\ \Eprint {http://arxiv.org/abs/hep-lat/0307017}
  {arXiv:hep-lat/0307017} \BibitemShut {NoStop}%
\bibitem [{\citenamefont {Attems}\ \emph {et~al.}(2017)\citenamefont {Attems},
  \citenamefont {Bea}, \citenamefont {Casalderrey-Solana}, \citenamefont
  {Mateos}, \citenamefont {Triana},\ and\ \citenamefont
  {Zilhao}}]{Attems:2017ezz}%
  \BibitemOpen
  \bibfield  {author} {\bibinfo {author} {\bibfnamefont {M.}~\bibnamefont
  {Attems}}, \bibinfo {author} {\bibfnamefont {Y.}~\bibnamefont {Bea}},
  \bibinfo {author} {\bibfnamefont {J.}~\bibnamefont {Casalderrey-Solana}},
  \bibinfo {author} {\bibfnamefont {D.}~\bibnamefont {Mateos}}, \bibinfo
  {author} {\bibfnamefont {M.}~\bibnamefont {Triana}}, \ and\ \bibinfo {author}
  {\bibfnamefont {M.}~\bibnamefont {Zilhao}},\ }\href {\doibase
  10.1007/JHEP06(2017)129} {\bibfield  {journal} {\bibinfo  {journal} {JHEP}\
  }\textbf {\bibinfo {volume} {06}},\ \bibinfo {pages} {129} (\bibinfo {year}
  {2017})},\ \Eprint {http://arxiv.org/abs/1703.02948} {arXiv:1703.02948
  [hep-th]} \BibitemShut {NoStop}%
\bibitem [{\citenamefont {Janik}\ \emph {et~al.}(2017)\citenamefont {Janik},
  \citenamefont {Jankowski},\ and\ \citenamefont
  {Soltanpanahi}}]{Janik:2017ykj}%
  \BibitemOpen
  \bibfield  {author} {\bibinfo {author} {\bibfnamefont {R.~A.}\ \bibnamefont
  {Janik}}, \bibinfo {author} {\bibfnamefont {J.}~\bibnamefont {Jankowski}}, \
  and\ \bibinfo {author} {\bibfnamefont {H.}~\bibnamefont {Soltanpanahi}},\
  }\href {\doibase 10.1103/PhysRevLett.119.261601} {\bibfield  {journal}
  {\bibinfo  {journal} {Phys. Rev. Lett.}\ }\textbf {\bibinfo {volume} {119}},\
  \bibinfo {pages} {261601} (\bibinfo {year} {2017})},\ \Eprint
  {http://arxiv.org/abs/1704.05387} {arXiv:1704.05387 [hep-th]} \BibitemShut
  {NoStop}%
\bibitem [{\citenamefont {Attems}\ \emph {et~al.}(2020)\citenamefont {Attems},
  \citenamefont {Bea}, \citenamefont {Casalderrey-Solana}, \citenamefont
  {Mateos},\ and\ \citenamefont {Zilh{\~a}o}}]{Attems:2019yqn}%
  \BibitemOpen
  \bibfield  {author} {\bibinfo {author} {\bibfnamefont {M.}~\bibnamefont
  {Attems}}, \bibinfo {author} {\bibfnamefont {Y.}~\bibnamefont {Bea}},
  \bibinfo {author} {\bibfnamefont {J.}~\bibnamefont {Casalderrey-Solana}},
  \bibinfo {author} {\bibfnamefont {D.}~\bibnamefont {Mateos}}, \ and\ \bibinfo
  {author} {\bibfnamefont {M.}~\bibnamefont {Zilh{\~a}o}},\ }\href {\doibase
  10.1007/JHEP01(2020)106} {\bibfield  {journal} {\bibinfo  {journal} {JHEP}\
  }\textbf {\bibinfo {volume} {01}},\ \bibinfo {pages} {106} (\bibinfo {year}
  {2020})},\ \Eprint {http://arxiv.org/abs/1905.12544} {arXiv:1905.12544
  [hep-th]} \BibitemShut {NoStop}%
\bibitem [{\citenamefont {Bellantuono}\ \emph {et~al.}(2019)\citenamefont
  {Bellantuono}, \citenamefont {Janik}, \citenamefont {Jankowski},\ and\
  \citenamefont {Soltanpanahi}}]{Bellantuono:2019wbn}%
  \BibitemOpen
  \bibfield  {author} {\bibinfo {author} {\bibfnamefont {L.}~\bibnamefont
  {Bellantuono}}, \bibinfo {author} {\bibfnamefont {R.~A.}\ \bibnamefont
  {Janik}}, \bibinfo {author} {\bibfnamefont {J.}~\bibnamefont {Jankowski}}, \
  and\ \bibinfo {author} {\bibfnamefont {H.}~\bibnamefont {Soltanpanahi}},\
  }\href {\doibase 10.1007/JHEP10(2019)146} {\bibfield  {journal} {\bibinfo
  {journal} {JHEP}\ }\textbf {\bibinfo {volume} {10}},\ \bibinfo {pages} {146}
  (\bibinfo {year} {2019})},\ \Eprint {http://arxiv.org/abs/1906.00061}
  {arXiv:1906.00061 [hep-th]} \BibitemShut {NoStop}%
\bibitem [{\citenamefont {Li}\ \emph {et~al.}(2020)\citenamefont {Li},
  \citenamefont {Nie},\ and\ \citenamefont {Tian}}]{Li:2020ayr}%
  \BibitemOpen
  \bibfield  {author} {\bibinfo {author} {\bibfnamefont {X.}~\bibnamefont
  {Li}}, \bibinfo {author} {\bibfnamefont {Z.-Y.}\ \bibnamefont {Nie}}, \ and\
  \bibinfo {author} {\bibfnamefont {Y.}~\bibnamefont {Tian}},\ }\href {\doibase
  10.1007/JHEP09(2020)063} {\bibfield  {journal} {\bibinfo  {journal} {JHEP}\
  }\textbf {\bibinfo {volume} {09}},\ \bibinfo {pages} {063} (\bibinfo {year}
  {2020})},\ \Eprint {http://arxiv.org/abs/2003.12987} {arXiv:2003.12987
  [hep-th]} \BibitemShut {NoStop}%
\bibitem [{\citenamefont {Bergman}\ \emph {et~al.}(2007)\citenamefont
  {Bergman}, \citenamefont {Lifschytz},\ and\ \citenamefont
  {Lippert}}]{Bergman:2007wp}%
  \BibitemOpen
  \bibfield  {author} {\bibinfo {author} {\bibfnamefont {O.}~\bibnamefont
  {Bergman}}, \bibinfo {author} {\bibfnamefont {G.}~\bibnamefont {Lifschytz}},
  \ and\ \bibinfo {author} {\bibfnamefont {M.}~\bibnamefont {Lippert}},\ }\href
  {\doibase 10.1088/1126-6708/2007/11/056} {\bibfield  {journal} {\bibinfo
  {journal} {JHEP}\ }\textbf {\bibinfo {volume} {11}},\ \bibinfo {pages} {056}
  (\bibinfo {year} {2007})},\ \Eprint {http://arxiv.org/abs/0708.0326}
  {arXiv:0708.0326 [hep-th]} \BibitemShut {NoStop}%
\bibitem [{\citenamefont {Jokela}\ \emph {et~al.}(2021)\citenamefont {Jokela},
  \citenamefont {J{\"a}rvinen}, \citenamefont {Nijs},\ and\ \citenamefont
  {Remes}}]{Jokela:2020piw}%
  \BibitemOpen
  \bibfield  {author} {\bibinfo {author} {\bibfnamefont {N.}~\bibnamefont
  {Jokela}}, \bibinfo {author} {\bibfnamefont {M.}~\bibnamefont
  {J{\"a}rvinen}}, \bibinfo {author} {\bibfnamefont {G.}~\bibnamefont {Nijs}},
  \ and\ \bibinfo {author} {\bibfnamefont {J.}~\bibnamefont {Remes}},\ }\href
  {\doibase 10.1103/PhysRevD.103.086004} {\bibfield  {journal} {\bibinfo
  {journal} {Phys. Rev. D}\ }\textbf {\bibinfo {volume} {103}},\ \bibinfo
  {pages} {086004} (\bibinfo {year} {2021})},\ \Eprint
  {http://arxiv.org/abs/2006.01141} {arXiv:2006.01141 [hep-ph]} \BibitemShut
  {NoStop}%
\bibitem [{\citenamefont {Kovensky}\ and\ \citenamefont
  {Schmitt}(2020)}]{Kovensky:2020xif}%
  \BibitemOpen
  \bibfield  {author} {\bibinfo {author} {\bibfnamefont {N.}~\bibnamefont
  {Kovensky}}\ and\ \bibinfo {author} {\bibfnamefont {A.}~\bibnamefont
  {Schmitt}},\ }\href {\doibase 10.1007/JHEP09(2020)112} {\bibfield  {journal}
  {\bibinfo  {journal} {JHEP}\ }\textbf {\bibinfo {volume} {09}},\ \bibinfo
  {pages} {112} (\bibinfo {year} {2020})},\ \Eprint
  {http://arxiv.org/abs/2006.13739} {arXiv:2006.13739 [hep-th]} \BibitemShut
  {NoStop}%
\bibitem [{\citenamefont {Mintz}\ \emph {et~al.}(2010)\citenamefont {Mintz},
  \citenamefont {Fraga}, \citenamefont {Pagliara},\ and\ \citenamefont
  {Schaffner-Bielich}}]{Mintz:2009ay}%
  \BibitemOpen
  \bibfield  {author} {\bibinfo {author} {\bibfnamefont {B.~W.}\ \bibnamefont
  {Mintz}}, \bibinfo {author} {\bibfnamefont {E.~S.}\ \bibnamefont {Fraga}},
  \bibinfo {author} {\bibfnamefont {G.}~\bibnamefont {Pagliara}}, \ and\
  \bibinfo {author} {\bibfnamefont {J.}~\bibnamefont {Schaffner-Bielich}},\
  }\href {\doibase 10.1103/PhysRevD.81.123012} {\bibfield  {journal} {\bibinfo
  {journal} {Phys. Rev. D}\ }\textbf {\bibinfo {volume} {81}},\ \bibinfo
  {pages} {123012} (\bibinfo {year} {2010})},\ \Eprint
  {http://arxiv.org/abs/0910.3927} {arXiv:0910.3927 [hep-ph]} \BibitemShut
  {NoStop}%
\bibitem [{\citenamefont {Sagert}\ \emph {et~al.}(2009)\citenamefont {Sagert},
  \citenamefont {Fischer}, \citenamefont {Hempel}, \citenamefont {Pagliara},
  \citenamefont {Schaffner-Bielich}, \citenamefont {Mezzacappa}, \citenamefont
  {Thielemann},\ and\ \citenamefont {Liebendorfer}}]{Sagert:2008ka}%
  \BibitemOpen
  \bibfield  {author} {\bibinfo {author} {\bibfnamefont {I.}~\bibnamefont
  {Sagert}}, \bibinfo {author} {\bibfnamefont {T.}~\bibnamefont {Fischer}},
  \bibinfo {author} {\bibfnamefont {M.}~\bibnamefont {Hempel}}, \bibinfo
  {author} {\bibfnamefont {G.}~\bibnamefont {Pagliara}}, \bibinfo {author}
  {\bibfnamefont {J.}~\bibnamefont {Schaffner-Bielich}}, \bibinfo {author}
  {\bibfnamefont {A.}~\bibnamefont {Mezzacappa}}, \bibinfo {author}
  {\bibfnamefont {F.~K.}\ \bibnamefont {Thielemann}}, \ and\ \bibinfo {author}
  {\bibfnamefont {M.}~\bibnamefont {Liebendorfer}},\ }\href {\doibase
  10.1103/PhysRevLett.102.081101} {\bibfield  {journal} {\bibinfo  {journal}
  {Phys. Rev. Lett.}\ }\textbf {\bibinfo {volume} {102}},\ \bibinfo {pages}
  {081101} (\bibinfo {year} {2009})},\ \Eprint {http://arxiv.org/abs/0809.4225}
  {arXiv:0809.4225 [astro-ph]} \BibitemShut {NoStop}%
\bibitem [{\citenamefont {Logoteta}\ \emph {et~al.}(2012)\citenamefont
  {Logoteta}, \citenamefont {Providencia}, \citenamefont {Vidana},\ and\
  \citenamefont {Bombaci}}]{Logoteta:2012ms}%
  \BibitemOpen
  \bibfield  {author} {\bibinfo {author} {\bibfnamefont {D.}~\bibnamefont
  {Logoteta}}, \bibinfo {author} {\bibfnamefont {C.}~\bibnamefont
  {Providencia}}, \bibinfo {author} {\bibfnamefont {I.}~\bibnamefont {Vidana}},
  \ and\ \bibinfo {author} {\bibfnamefont {I.}~\bibnamefont {Bombaci}},\ }\href
  {\doibase 10.1103/PhysRevC.85.055807} {\bibfield  {journal} {\bibinfo
  {journal} {Phys. Rev. C}\ }\textbf {\bibinfo {volume} {85}},\ \bibinfo
  {pages} {055807} (\bibinfo {year} {2012})},\ \Eprint
  {http://arxiv.org/abs/1204.5909} {arXiv:1204.5909 [nucl-th]} \BibitemShut
  {NoStop}%
\bibitem [{\citenamefont {Most}\ \emph {et~al.}(2018)\citenamefont {Most},
  \citenamefont {Weih}, \citenamefont {Rezzolla},\ and\ \citenamefont
  {Schaffner-Bielich}}]{Most:2018hfd}%
  \BibitemOpen
  \bibfield  {author} {\bibinfo {author} {\bibfnamefont {E.~R.}\ \bibnamefont
  {Most}}, \bibinfo {author} {\bibfnamefont {L.~R.}\ \bibnamefont {Weih}},
  \bibinfo {author} {\bibfnamefont {L.}~\bibnamefont {Rezzolla}}, \ and\
  \bibinfo {author} {\bibfnamefont {J.}~\bibnamefont {Schaffner-Bielich}},\
  }\href {\doibase 10.1103/PhysRevLett.120.261103} {\bibfield  {journal}
  {\bibinfo  {journal} {Phys. Rev. Lett.}\ }\textbf {\bibinfo {volume} {120}},\
  \bibinfo {pages} {261103} (\bibinfo {year} {2018})},\ \Eprint
  {http://arxiv.org/abs/1803.00549} {arXiv:1803.00549 [gr-qc]} \BibitemShut
  {NoStop}%
\bibitem [{\citenamefont {Bauswein}\ \emph {et~al.}(2019)\citenamefont
  {Bauswein}, \citenamefont {Bastian}, \citenamefont {Blaschke}, \citenamefont
  {Chatziioannou}, \citenamefont {Clark}, \citenamefont {Fischer},\ and\
  \citenamefont {Oertel}}]{Bauswein:2018bma}%
  \BibitemOpen
  \bibfield  {author} {\bibinfo {author} {\bibfnamefont {A.}~\bibnamefont
  {Bauswein}}, \bibinfo {author} {\bibfnamefont {N.-U.~F.}\ \bibnamefont
  {Bastian}}, \bibinfo {author} {\bibfnamefont {D.~B.}\ \bibnamefont
  {Blaschke}}, \bibinfo {author} {\bibfnamefont {K.}~\bibnamefont
  {Chatziioannou}}, \bibinfo {author} {\bibfnamefont {J.~A.}\ \bibnamefont
  {Clark}}, \bibinfo {author} {\bibfnamefont {T.}~\bibnamefont {Fischer}}, \
  and\ \bibinfo {author} {\bibfnamefont {M.}~\bibnamefont {Oertel}},\ }\href
  {\doibase 10.1103/PhysRevLett.122.061102} {\bibfield  {journal} {\bibinfo
  {journal} {Phys. Rev. Lett.}\ }\textbf {\bibinfo {volume} {122}},\ \bibinfo
  {pages} {061102} (\bibinfo {year} {2019})},\ \Eprint
  {http://arxiv.org/abs/1809.01116} {arXiv:1809.01116 [astro-ph.HE]}
  \BibitemShut {NoStop}%
\bibitem [{\citenamefont {Chesler}\ \emph {et~al.}(2019)\citenamefont
  {Chesler}, \citenamefont {Jokela}, \citenamefont {Loeb},\ and\ \citenamefont
  {Vuorinen}}]{Chesler:2019osn}%
  \BibitemOpen
  \bibfield  {author} {\bibinfo {author} {\bibfnamefont {P.~M.}\ \bibnamefont
  {Chesler}}, \bibinfo {author} {\bibfnamefont {N.}~\bibnamefont {Jokela}},
  \bibinfo {author} {\bibfnamefont {A.}~\bibnamefont {Loeb}}, \ and\ \bibinfo
  {author} {\bibfnamefont {A.}~\bibnamefont {Vuorinen}},\ }\href {\doibase
  10.1103/PhysRevD.100.066027} {\bibfield  {journal} {\bibinfo  {journal}
  {Phys. Rev. D}\ }\textbf {\bibinfo {volume} {100}},\ \bibinfo {pages}
  {066027} (\bibinfo {year} {2019})},\ \Eprint
  {http://arxiv.org/abs/1906.08440} {arXiv:1906.08440 [astro-ph.HE]}
  \BibitemShut {NoStop}%
\bibitem [{\citenamefont {Glendenning}(1992)}]{Glendenning:1992vb}%
  \BibitemOpen
  \bibfield  {author} {\bibinfo {author} {\bibfnamefont {N.~K.}\ \bibnamefont
  {Glendenning}},\ }\href {\doibase 10.1103/PhysRevD.46.1274} {\bibfield
  {journal} {\bibinfo  {journal} {Phys. Rev. D}\ }\textbf {\bibinfo {volume}
  {46}},\ \bibinfo {pages} {1274} (\bibinfo {year} {1992})}\BibitemShut
  {NoStop}%
\bibitem [{\citenamefont {Heiselberg}\ \emph {et~al.}(1993)\citenamefont
  {Heiselberg}, \citenamefont {Pethick},\ and\ \citenamefont
  {Staubo}}]{Heiselberg:1992dx}%
  \BibitemOpen
  \bibfield  {author} {\bibinfo {author} {\bibfnamefont {H.}~\bibnamefont
  {Heiselberg}}, \bibinfo {author} {\bibfnamefont {C.~J.}\ \bibnamefont
  {Pethick}}, \ and\ \bibinfo {author} {\bibfnamefont {E.~F.}\ \bibnamefont
  {Staubo}},\ }\href {\doibase 10.1103/PhysRevLett.70.1355} {\bibfield
  {journal} {\bibinfo  {journal} {Phys. Rev. Lett.}\ }\textbf {\bibinfo
  {volume} {70}},\ \bibinfo {pages} {1355} (\bibinfo {year}
  {1993})}\BibitemShut {NoStop}%
\bibitem [{\citenamefont {Kurkela}\ \emph {et~al.}(2014)\citenamefont
  {Kurkela}, \citenamefont {Fraga}, \citenamefont {Schaffner-Bielich},\ and\
  \citenamefont {Vuorinen}}]{Kurkela:2014vha}%
  \BibitemOpen
  \bibfield  {author} {\bibinfo {author} {\bibfnamefont {A.}~\bibnamefont
  {Kurkela}}, \bibinfo {author} {\bibfnamefont {E.~S.}\ \bibnamefont {Fraga}},
  \bibinfo {author} {\bibfnamefont {J.}~\bibnamefont {Schaffner-Bielich}}, \
  and\ \bibinfo {author} {\bibfnamefont {A.}~\bibnamefont {Vuorinen}},\ }\href
  {\doibase 10.1088/0004-637X/789/2/127} {\bibfield  {journal} {\bibinfo
  {journal} {Astrophys. J.}\ }\textbf {\bibinfo {volume} {789}},\ \bibinfo
  {pages} {127} (\bibinfo {year} {2014})},\ \Eprint
  {http://arxiv.org/abs/1402.6618} {arXiv:1402.6618 [astro-ph.HE]} \BibitemShut
  {NoStop}%
\bibitem [{\citenamefont {Glendenning}(1997)}]{Glendenning:1997wn}%
  \BibitemOpen
  \bibfield  {author} {\bibinfo {author} {\bibfnamefont {N.~K.}\ \bibnamefont
  {Glendenning}},\ }\href@noop {} {\emph {\bibinfo {title} {{Compact stars:
  Nuclear physics, particle physics, and general relativity}}}}\ (\bibinfo
  {year} {1997})\BibitemShut {NoStop}%
\bibitem [{\citenamefont {Palhares}\ and\ \citenamefont
  {Fraga}(2010)}]{Palhares:2010be}%
  \BibitemOpen
  \bibfield  {author} {\bibinfo {author} {\bibfnamefont {L.~F.}\ \bibnamefont
  {Palhares}}\ and\ \bibinfo {author} {\bibfnamefont {E.~S.}\ \bibnamefont
  {Fraga}},\ }\href {\doibase 10.1103/PhysRevD.82.125018} {\bibfield  {journal}
  {\bibinfo  {journal} {Phys. Rev. D}\ }\textbf {\bibinfo {volume} {82}},\
  \bibinfo {pages} {125018} (\bibinfo {year} {2010})},\ \Eprint
  {http://arxiv.org/abs/1006.2357} {arXiv:1006.2357 [hep-ph]} \BibitemShut
  {NoStop}%
\bibitem [{\citenamefont {Fraga}\ \emph {et~al.}(2019)\citenamefont {Fraga},
  \citenamefont {Hippert},\ and\ \citenamefont {Schmitt}}]{Fraga:2018cvr}%
  \BibitemOpen
  \bibfield  {author} {\bibinfo {author} {\bibfnamefont {E.~S.}\ \bibnamefont
  {Fraga}}, \bibinfo {author} {\bibfnamefont {M.}~\bibnamefont {Hippert}}, \
  and\ \bibinfo {author} {\bibfnamefont {A.}~\bibnamefont {Schmitt}},\ }\href
  {\doibase 10.1103/PhysRevD.99.014046} {\bibfield  {journal} {\bibinfo
  {journal} {Phys. Rev. D}\ }\textbf {\bibinfo {volume} {99}},\ \bibinfo
  {pages} {014046} (\bibinfo {year} {2019})},\ \Eprint
  {http://arxiv.org/abs/1810.13226} {arXiv:1810.13226 [hep-ph]} \BibitemShut
  {NoStop}%
\end{thebibliography}%


%

\end{document}